\documentclass[twocolumn,tighten]{aastex61}
\pdfoutput=1 
\usepackage{amsmath,amstext}
\usepackage[T1]{fontenc}
\usepackage{apjfonts} 
\usepackage[figure,figure*]{hypcap}

\usepackage{hyperref}

\usepackage{etoolbox}

\makeatletter

\patchcmd{\NAT@citex}
{\@citea\NAT@hyper@{%
\NAT@nmfmt{\NAT@nm}%
\hyper@natlinkbreak{\NAT@aysep\NAT@spacechar}{\@citeb\@extra@b@citeb}%
\NAT@date}}
{\@citea\NAT@nmfmt{\NAT@nm}%
\NAT@aysep\NAT@spacechar\NAT@hyper@{\NAT@date}}{}{}

\patchcmd{\NAT@citex}
{\@citea\NAT@hyper@{%
\NAT@nmfmt{\NAT@nm}%
\hyper@natlinkbreak{\NAT@spacechar\NAT@@open\if#1\else#1\NAT@spacechar\fi}%
{\@citeb\@extra@b@citeb}%
\NAT@date}}
{\@citea\NAT@nmfmt{\NAT@nm}%
\NAT@spacechar\NAT@@open\if#1\else#1\NAT@spacechar\fi\NAT@hyper@{\NAT@date}}
{}{}

\makeatother

\newcommand*{\http}[1]{\href{http://#1}{#1}}

\shorttitle{Predicting Realistic Subhalo Populations}
\shortauthors{Nadler et al.}

\begin{document}

\title{Modeling the Impact of Baryons on Subhalo Populations with Machine Learning}

\author[0000-0002-1182-3825]{Ethan~O.~Nadler}
\affiliation{Kavli Institute for Particle Astrophysics and Cosmology and Department of Physics, Stanford University, Stanford, CA 94305, USA}
\author[0000-0002-1200-0820]{Yao-Yuan~Mao}
\affiliation{Department of Physics and Astronomy and Pittsburgh Particle Physics, Astrophysics and Cosmology Center (PITT PACC), University of Pittsburgh, Pittsburgh, PA 15260, USA}
\author[0000-0003-2229-011X]{Risa~H.~Wechsler}
\affiliation{Kavli Institute for Particle Astrophysics and Cosmology and Department of Physics, Stanford University, Stanford, CA 94305, USA}
\affiliation{SLAC National Accelerator Laboratory, Menlo Park, CA 94025, USA}
\author[0000-0002-4655-8128]{Shea~Garrison-Kimmel}
\affiliation{TAPIR, Mailcode 350-17, California Institute of Technology, Pasadena, CA 91125, USA}
\author[0000-0003-0603-8942]{Andrew~Wetzel}
\affiliation{TAPIR, Mailcode 350-17, California Institute of Technology, Pasadena, CA 91125, USA}
\affiliation{The Observatories of the Carnegie Institution for Science, Pasadena, CA 91101, USA}
\affiliation{Department of Physics, University of California, Davis, CA 95616, USA}

\correspondingauthor{Ethan~O.~Nadler}
\email{enadler@stanford.edu}


\begin{abstract}
We identify subhalos in dark matter-only (DMO) zoom-in simulations that are likely to be disrupted due to baryonic effects by using a random forest classifier trained on two hydrodynamic simulations of Milky Way (MW)-mass host halos from the Latte suite of the Feedback in Realistic Environments (FIRE) project. We train our classifier using five properties of each disrupted and surviving subhalo: pericentric distance and scale factor at first pericentric passage after accretion, and scale factor, virial mass, and maximum circular velocity at accretion. Our five-property classifier identifies disrupted subhalos in the FIRE simulations with an $85\%$ out-of-bag classification score. We predict surviving subhalo populations in DMO simulations of the FIRE host halos, finding excellent agreement with the hydrodynamic results; in particular, our classifier outperforms DMO zoom-in simulations that include the gravitational potential of the central galactic disk in each hydrodynamic simulation, indicating that it captures both the dynamical effects of a central disk and additional baryonic physics. We also predict surviving subhalo populations for a suite of DMO zoom-in simulations of MW-mass host halos, finding that baryons impact each system consistently and that the predicted amount of subhalo disruption is larger than the host-to-host scatter among the subhalo populations. Although the small size and specific baryonic physics prescription of our training set limits the generality of our results, our work suggests that machine-learning classification algorithms trained on hydrodynamic zoom-in simulations can efficiently predict realistic subhalo populations.
\end{abstract}



\section{Introduction}
\label{Introduction}

The $\Lambda$CDM cosmological model provides a remarkably successful framework in which the observed large-scale distribution of galaxies can be understood in terms of the underlying distribution of dark matter halos. However, there are several outstanding ``small-scale'' problems associated with $\Lambda$CDM cosmology (see \citealp{Bullock170704256} and \citealp{DelPopolo160607790} for recent reviews). For example, dark matter-only (DMO) simulations predict large numbers of low-mass subhalos that contribute to an ever-rising low-mass end of the subhalo mass function. If these low-mass subhalos exist and host galaxies, we should observe many more dwarf satellites than currently detected around the Milky Way (MW) or the Andromeda Galaxy (M31); this is often dubbed the missing-satellites problem \citep{Klypin9901240,Moore9907411}. Meanwhile, the ``too big to fail'' \citep[TBTF;][]{Boylan-Kolchin11030007} problem arises because the number of subhalos with high maximum circular velocities ($V_{\rm max}\gtrsim 15\ \rm{km\ s}^{-1}$) found in DMO simulations of MW-mass systems substantially exceeds the number of such subhalos inferred to exist around the MW and M31. Equivalently, observational estimates for the masses of the subhalos that host the dwarf satellites of the MW and M31 fall below the masses predicted by DMO simulations \citep{Boylan-Kolchin11112048}.

While these small-scale problems present challenges to the $\Lambda$CDM paradigm, a number of promising astrophysical solutions to each problem have been proposed. For example, it is now understood that cosmic reionization suppresses star formation in low-mass subhalos, while supernova (SN) feedback can suppress star formation in more massive subhalos, potentially resolving the missing-satellites problem \citep{Bullock0002214,Somerville0107507}. Proposed solutions to the TBTF problem build on these ideas by invoking stellar feedback to soften central density cusps and deplete subhalos of dark matter \citep{Governato12020554,Pontzen11060499}, along with enhanced subhalo disruption via tidal stripping or disk shocking, to destroy many of the high-$V_{\rm max}$ subhalos found in DMO simulations. Several authors have suggested that these mechanisms can yield subhalo populations in agreement with those inferred observationally for the MW and M31 \citep{Zolotov12070007,Brooks12095394,Brook14103825,Wetzel160205957,Sawala160901718}.

Indeed, recent high-resolution hydrodynamic simulations that self-consistently resolve star formation, stellar feedback, and the formation of central galactic disks indicate that the missing-satellites and TBTF problems can largely be mitigated for the subhalo populations of MW-mass host halos. For example, Wetzel et al.\ (\citeyear{Wetzel160205957}) and\defcitealias{Garrison-Kimmel170103792}{GK17} Garrison-Kimmel et al.\ (\citeyear{Garrison-Kimmel170103792}, hereafter \citetalias{Garrison-Kimmel170103792}) studied the subhalo populations of two MW-mass host halos from the Latte simulation suite of the Feedback in Realistic Environments (FIRE) project \citep{Hopkins13112073} using the `zoom-in' simulation technique \citep{Katz172935,Onorbe150202036}. These authors found that the total number of subhalos in each simulation is reduced by about a factor of two relative to corresponding DMO simulations with identical initial conditions, and they also found significantly fewer subhalos with high circular velocities in the hydrodynamic runs. Moreover, the subhalo populations in both of these systems are consistent with a variety of observational probes for the MW and M31, which suggests that the missing-satellites and TBTF problems can be resolved in these particular simulations \citep{Wetzel160205957}. Zhu et al.\ (\citeyear{Zhu150605537}) reached similar conclusions by comparing hydrodynamic zoom-in simulations of MW-mass host halos from the Aquarius Project \citep{Springel08090898} to DMO simulations of the same hosts.

These results rely on a limited number of high-resolution simulations of MW-mass host halos; unfortunately, studying a large, diverse sample of subhalo populations in hydrodynamic zoom-in simulations is currently infeasible. While many authors have justifiably focused on the subhalo populations of MW-mass host halos, since these are particularly relevant to the original TBTF problem, it is important to assess whether analogous TBTF problems arise for the subhalo populations of more massive host halos. In addition, understanding whether the TBTF problem is consistently mitigated in a range of simulations with different baryonic physics implementations is necessary in order to make robust conclusions. Quantifying the impact of baryonic physics on subhalo populations more generally will be important in order to interpret results from large-scale surveys, including the Dark Energy Spectroscopic Instrument \citep{DESI161100036} and the Large Synoptic Survey Telescope \citep{LSST09120201}, and from targeted searches for satellites of MW-like galaxies outside the Local Group such as the Satellites Around Galactic Analogs Survey \citep[SAGA;][]{Mao170506743}.

Thus, models that can incorporate a variety of hydrodynamic simulations to predict realistic subhalo populations directly from DMO simulations are worth exploring. As a first step toward such a model, we present a machine-learning classification algorithm to identify subhalos in DMO zoom-in simulations of MW-mass host halos that are likely to be disrupted due to baryonic effects in hydrodynamic resimulations. In particular, we train a random forest classifier on disrupted and surviving subhalos from the FIRE zoom-in simulations presented in \citetalias{Garrison-Kimmel170103792}, and we use the classifier to predict surviving subhalo populations in DMO zoom-in simulations. Our aim is to explore whether this algorithm can capture the effects of baryons in existing hydrodynamic simulations and how the particular baryonic physics in these simulations alters subhalo populations in independent DMO simulations. Rather than providing a detailed comparison of different classification algorithms, we show that a simple random forest classifier predicts subhalo populations in excellent agreement with hydrodynamic results when applied to DMO simulations of the FIRE host halos. This technique is efficient, since a trained classifier can immediately predict surviving subhalo populations from relatively inexpensive DMO simulations. We view classification as a promising technique for predicting subhalo disruption because classifiers will become more robust as the number of high-resolution hydrodynamic simulations to train on increases. In particular, classification algorithms can be trained on a variety of zoom-in simulations to capture the impact of baryons on subhalo populations for a range of host halo masses, central galaxy types, formation histories, and subgrid physics prescriptions.

In addition to the practical utility of our results for predicting realistic subhalo populations, our work provides insights into subhalo disruption in hydrodynamic simulations and relates to the small-scale challenges described above. For example, our random forest classifier determines how strongly various subhalo properties correlate with disruption, which indicates the importance of different disruption mechanisms, including tidal effects and stellar feedback, given the specific baryonic physics prescription in these simulations. To explore the relative importance of these disruption mechanisms, we compare the surviving subhalo populations that we predict from DMO simulations of the FIRE host halos to the DMO-plus-disk simulations presented in \citetalias{Garrison-Kimmel170103792}, which are designed to capture the dynamical effects of the central galactic disk that develops in each hydrodynamic simulation. In particular, by performing DMO zoom-in simulations of two systems with analytic disk potentials tuned to match the galactic disks that develop in the corresponding hydrodynamic simulations, \citetalias{Garrison-Kimmel170103792} found subhalo populations in good agreement with the hydrodynamic results, particularly in the innermost regions ($r\lesssim 100\ \rm{kpc}$). This result suggests that, for MW-mass halos with a central galactic disk, the tidal effects of the disk are largely responsible for disrupting both the low-$V_{\rm max}$ subhalos relevant to the missing-satellites problem and the high-$V_{\rm max}$ subhalos relevant to the TBTF problem. Our machine-learning predictions are consistent with the DMO-plus-disk simulations at low $V_{\rm max}$, but we find enhanced disruption for subhalos with $V_{\rm max}\gtrsim 15\ \rm{km\ s}^{-1}$ and our results match the FIRE simulations more closely for such subhalos. Interestingly, several authors have suggested that baryonic physics efficiently creates cored subhalo density profiles in this regime \citep{Chan150702282,Tollet150703590,Fitts161102281}. We therefore argue that baryonic effects within subhalos, such as stellar feedback, can help to relieve the tension between the subhalo populations predicted by DMO simulations and those inferred from observations of the Local Group.

Our work also has broader implications for studying the galaxy--halo connection. For example, by using our classifier to predict surviving subhalo populations for the suite of DMO zoom-in simulations of MW-mass host halos from Mao et al.\ (\citeyear{Mao150302637}), we find that the average amount of subhalo disruption due to baryonic effects is larger than the host-to-host scatter among various subhalo populations. Thus, models that utilize subhalo statistics from these simulations should account for enhanced subhalo disruption when marginalizing over the effects of baryonic physics. Several semianalytic models (e.g.,~\citealp{Lu160502075,Lu170307467}) incorporate subhalo velocity functions predicted by DMO zoom-in simulations of MW-mass host halos in order to constrain the properties of the MW satellite galaxies and their host halos, and it is plausible that the physical insights provided by these models could change when more realistic subhalo populations are used as input.

This paper is organized as follows. In Section \ref{Data}, we describe the FIRE simulations that we use to train our random forest classifier, as well as the DMO and DMO-plus-disk simulations presented in \citetalias{Garrison-Kimmel170103792} to which we compare our results. In Section \ref{RandomForest}, we describe our training and cross-validation methods, and we test our classifier by predicting disrupted subhalos in two FIRE zoom-in simulations. We present our main results in Section \ref{Results}. In Section \ref{FIREML}, we predict surviving subhalo populations in DMO simulations of the FIRE host halos, and we present velocity functions and radial distributions for our predicted subhalo populations; in Section \ref{MWZoomin}, we predict surviving subhalo populations for the suite of DMO zoom-in simulations from Mao et al.\ (\citeyear{Mao150302637}), and we discuss the implications for satellite searches. We address avenues for future work and summarize our conclusions in Section \ref{Discussion}.

We adopt cosmological parameters consistent with each simulation that we analyze. In particular, we use $h = 0.702$, $\Omega_{\rm m} = 0.272$, $\Omega_{\rm b} = 0.0455$, and $\Omega_{\Lambda} = 0.728$ for our analysis of the FIRE simulations and $h = 0.7$, $\Omega_{\rm m} = 0.286$, $\Omega_{\rm b} = 0.047$, and $\Omega_{\Lambda} = 0.714$ for our analysis of the MW zoom-in simulation suite. Note that we express distances in physical $\rm{kpc}$ and velocities in $\rm{km\ s}^{-1}$.


\section{Simulation Data}
\label{Data}

We train our random forest classifier using subhalos from the hydrodynamic zoom-in simulations presented in \citetalias{Garrison-Kimmel170103792}. These authors studied the subhalo populations of two MW-mass host halos, referred to as \texttt{m12i} ($M_{\rm vir} = 1.1 \times 10^{12}\ M_{\rm \odot}$) and \texttt{m12f} ($M_{\rm vir} = 1.6 \times 10^{12}\ M_{\rm \odot}$), which were simulated as part of the Latte suite from the FIRE project \citep{Hopkins13112073}. These simulations were performed using the FIRE-2 code \citep{Hopkins170206148}, which includes the same radiative heating and cooling, star formation, and stellar feedback prescriptions as the original FIRE-1 code in addition to several numerical improvements. The simulations were run in the same cosmological volume (side length $60\ h^{-1} \ \rm{Mpc}$) as the AGORA project \citep{Kim13082669}; the \texttt{m12i} and \texttt{m12f} zoom-in simulation regions each contain a single host halo at redshift $z=0$ that has no MW-mass neighbors within $3\ \rm{Mpc}$. The \texttt{m12i} simulation was originally presented in Wetzel et al.\ (\citeyear{Wetzel160205957}); \texttt{m12f}, which was simulated using the same parameters and pipeline, was first presented in \citetalias{Garrison-Kimmel170103792}. The baryonic mass resolution in these simulations is $\sim 7000\ M_{\rm \odot}$, while the dark matter particle mass is $3.5\times 10^{4}\ M_{\rm \odot}$, corresponding to a subhalo mass resolution of $\sim 3\times 10^6\ M_{\rm \odot}$. We refer the reader to \citetalias{Garrison-Kimmel170103792} and Hopkins et al.\ (\citeyear{Hopkins170206148}) for details on the initial conditions, gravitational-force softenings, and models for radiative heating/cooling, star formation, and stellar feedback in these simulations. Halo catalogs were created using \texttt{AHF} \citep{Knollmann09043662} and merger trees were generated using the \texttt{consistent-trees} merger code \citep{Behroozi11104370}.

We will compare our results to the \texttt{m12i} and \texttt{m12f} subhalo populations from three sets of simulations: the hydrodynamic FIRE simulations described above, DMO simulations that were run with identical initial conditions, and the dark matter-plus-disk potential (DISK) simulations presented in \citetalias{Garrison-Kimmel170103792}. The DISK simulations are identical to the corresponding DMO simulations, but they include gravitational potentials designed to capture the effects of the central disks in the hydrodynamic simulations. In particular, a disk potential is added to each DMO zoom-in simulation at $z=3$, and its parameters and evolution are tuned to match the central disk that develops in the corresponding FIRE simulation. We refer the reader to \citetalias{Garrison-Kimmel170103792} for a detailed description of the DISK simulations.

Figure $1$ in \citetalias{Garrison-Kimmel170103792} illustrates the dark matter substructure in \texttt{m12i} for each type of simulation. The visual differences between the FIRE and DMO subhalo populations qualitatively show that baryonic physics in the FIRE simulations lowers both the total number of surviving subhalos and the number of high-$V_{\text{max}}$ subhalos that contribute to the TBTF problem. This figure also shows that the DISK simulation captures the majority of the subhalo disruption in \texttt{m12i}, particularly in the innermost regions ($r\lesssim 100\ \rm{kpc}$), which implies that the central disk is largely responsible for the subhalo disruption in the corresponding hydrodynamic simulation. We have verified the quantitative results in \citetalias{Garrison-Kimmel170103792} by calculating velocity functions and radial distributions for the \texttt{m12i} and \texttt{m12f} subhalo populations in the FIRE, DISK, and DMO simulations. Note that, as in \citetalias{Garrison-Kimmel170103792}, we scale all subhalo masses by a factor of $1-f_{\rm b}$ and all subhalo circular velocities by a factor of $\sqrt{1-f_{\rm b}}$ in our post-processing of the DMO and DISK halo catalogs, where $f_{\rm b} = \Omega_{\rm b}/\Omega_{\rm m} \simeq 0.17$ is the cosmic baryon fraction. The mass correction accounts for the fact that the baryonic mass in the hydrodynamic simulations is included in the dark matter particles in the DMO simulations, and the circular velocity correction is an approximate way to account for reduced subhalo densities due to stellar feedback, similar to the prescription in Zolotov et al.\ (\citeyear{Zolotov12070007}). Neither of these corrections affect our results.

To study disrupted subhalos in the FIRE simulations, we select subhalos that disappear from the \texttt{m12i} and \texttt{m12f} halo catalogs after $z=3$. We choose this redshift in order to match the initial redshift of the DISK simulations in \citetalias{Garrison-Kimmel170103792}; note that there are very few subhalos disrupted before $z=3$ that pass our subsequent minimum circular velocity cuts. We restrict our analysis to first-order subhalos (i.e., we exclude subhalos of subhalos); thus, for a disrupted subhalo to be included in our catalog, it must contribute to the host halo at $z=0$. Operationally, each disrupted subhalo must have a descendant ID equal to the ID of a main-branch progenitor of the final host halo. Meanwhile, we define surviving subhalos as those that remain in the halo catalog at $z=0$ and have a parent ID that is equal to the host ID, which similarly excludes higher-order subhalos.

To ensure that we study well-resolved subhalos, we restrict both disrupted and surviving subhalos to those with peak circular velocity $V_{\rm peak}>10\ \rm{km\ s}^{-1}$ in our fiducial model, where $V_{\rm peak}$ is defined as the largest maximum circular velocity a subhalo attains along its entire main branch. This is a conservative choice; for example, \citetalias{Garrison-Kimmel170103792} presented velocity functions using the cut $V_{\rm max}>5\ \rm{km\ s}^{-1}$, where $V_{\rm max}$ is the maximum circular velocity at $z=0$. However, this cut ensures that we train our algorithm on subhalos that are consistent with those we will classify in a lower-resolution zoom-in simulation suite. By choosing a $V_{\rm peak}$ threshold rather than a $V_{\rm max}$ threshold, we also avoid biasing our subhalo selection with a redshift-dependent cut, since $V_{\rm peak}$ --- unlike $V_{\rm max}$ --- is not defined at a particular redshift. The $V_{\rm peak}>10\ \rm{km\ s}^{-1}$ cut results in a combined total of $566$ surviving subhalos and $872$ disrupted subhalos from \texttt{m12i} and \texttt{m12f}, which we combine to form our fiducial training set. In Appendix \ref{appendix}, we examine the impact of different training sets and minimum circular velocity cuts, and we present the results using the $V_{\rm max}$ cut employed in \citetalias{Garrison-Kimmel170103792} for comparison.


\section{Random Forest Classification}
\label{RandomForest}

\subsection{Overview}

We use the random forest algorithm from the package \texttt{Scikit-Learn} \citep{scikit-learn} to classify disrupted and surviving subhalos. We refer the reader to the \texttt{Scikit-Learn} documentation for a detailed description of the algorithm, but we outline the most important aspects here. A random forest is a collection of decision trees, each of which is tuned to classify objects based on their input properties. Each tree in the forest is trained on a random sample of the training data with replacement, using a random subset of the input features at each split in the learning process, with the goal of predicting the classes of the objects in the training set as accurately as possible according to some metric. For example, the default \texttt{Scikit-Learn} implementation minimizes the Gini impurity of the classifier's prediction. The random forest prediction for a given object is the majority vote of the tuned decision trees, while the classification probability is equal to the fraction of trees that predict a certain class. In this work, we label subhalos as either surviving until $z=0$ or disrupted at some earlier time; thus, our random forest objects are subhalos, and our decision trees vote for whether each subhalo is disrupted or survives until $z=0$. Note that our model does not explicitly include enhanced mass stripping due to baryonic effects, since we simply label subhalos as disrupted or surviving.

We train our classifier using the disrupted and surviving subhalos from \texttt{m12i} and \texttt{m12f} described above. We train on subhalo properties that depend on the entire history of each subhalo to avoid biasing the classifier by using properties defined at specific redshifts --- for example, at $z=0$ for surviving subhalos or at the final available redshift for disrupted subhalos. In particular, since we aim to classify subhalos in DMO halo catalogs that have survived to $z=0$ but are likely to be disrupted in hydrodynamic resimulations, training our classifier with only present-day properties results in too many surviving subhalos because of the systematic evolution of subhalo properties over time.

Thus, we train on the following properties: pericentric distance and scale factor at first pericentric passage after accretion ($d_{\rm peri}$, $a_{\rm peri}$), and scale factor, virial mass, and maximum circular velocity at accretion ($a_{\rm acc}$, $M_{\rm acc}$, $V_{\rm acc}$). In principle, we could train the classifier on additional subhalo properties at pericenter or accretion; these properties could also include information about the host halo, such as subhalo scale radius in units of the host halo's scale radius. Indeed, random forests are well-suited to classifying objects using a large number of features because of the randomized nature of the training process, so we could even use \emph{every} available subhalo property at pericenter and accretion to train the classifier. However, we will show that our five-property classifier performs very well, so we adopt this model to simplify our analysis and avoid overfitting the training data. In addition, we checked whether including the present-day properties $V_{\text{max}}$ and $M_{\text{vir}}$ improves our classifier, finding that these properties are much less informative than features defined at pericenter or accretion. We discuss the correlations among the training features below, and we explore the feature selection in more detail in Appendix \ref{appendix}.

We calculate the aforementioned subhalo features from the merger trees as follows. We define accretion as the last snapshot, working backward in time from $z=0$ (for surviving subhalos) or from the redshift of disruption (for disrupted subhalos), at which a subhalo's host ID is equal to the main halo's ID. Physically, this occurs when a subhalo enters the virial radius of the host halo for the final time.\footnote{Note that a subhalo could have been contained within the host halo's virial radius at an earlier time and later reaccreted; we select the final accretion event for each subhalo.} We then take $a_{\rm acc}$, $M_{\rm acc}$, and $V_{\rm acc}$ as the scale factor, virial mass, and maximum circular velocity at the time of accretion for each subhalo. We define pericenter as the first snapshot after accretion at which a subhalo reaches a local minimum in its three-dimensional distance from the center of the host halo. We inspected individual subhalo orbits and determined that selecting the distance from the center of the host halo at the first snapshot after accretion at which a subhalo's separation from the host increases provides an accurate estimate of $d_{\rm peri}.$\footnote{Given a spacing of $\sim 25\ \rm{Myr}$ between halo catalog snapshots and a generous subhalo orbital velocity of $\sim 300\ \rm{km\ s}^{-1}$ at pericenter, the uncertainty in $d_{\rm peri}$ is only $\sim 8\ \rm{kpc}$.} For subhalos that do not reach a local minimum in their separation from the host halo after accretion, we define $d_{\rm peri}$ as the instantaneous distance from the center of the host. In particular, for surviving subhalos on infalling orbits that have not experienced a pericentric passage by $z=0$, we define $d_{\rm peri}$ as the distance from the host at $z=0$. Analogously, for destroyed subhalos on infalling orbits that have not reached pericenter by the time of disruption, we define $d_{\rm peri}$ as the distance from the host at the time of disruption.

\subsection{Choice of Subhalo Features}

\begin{figure}
\centering
\includegraphics[scale=0.43]{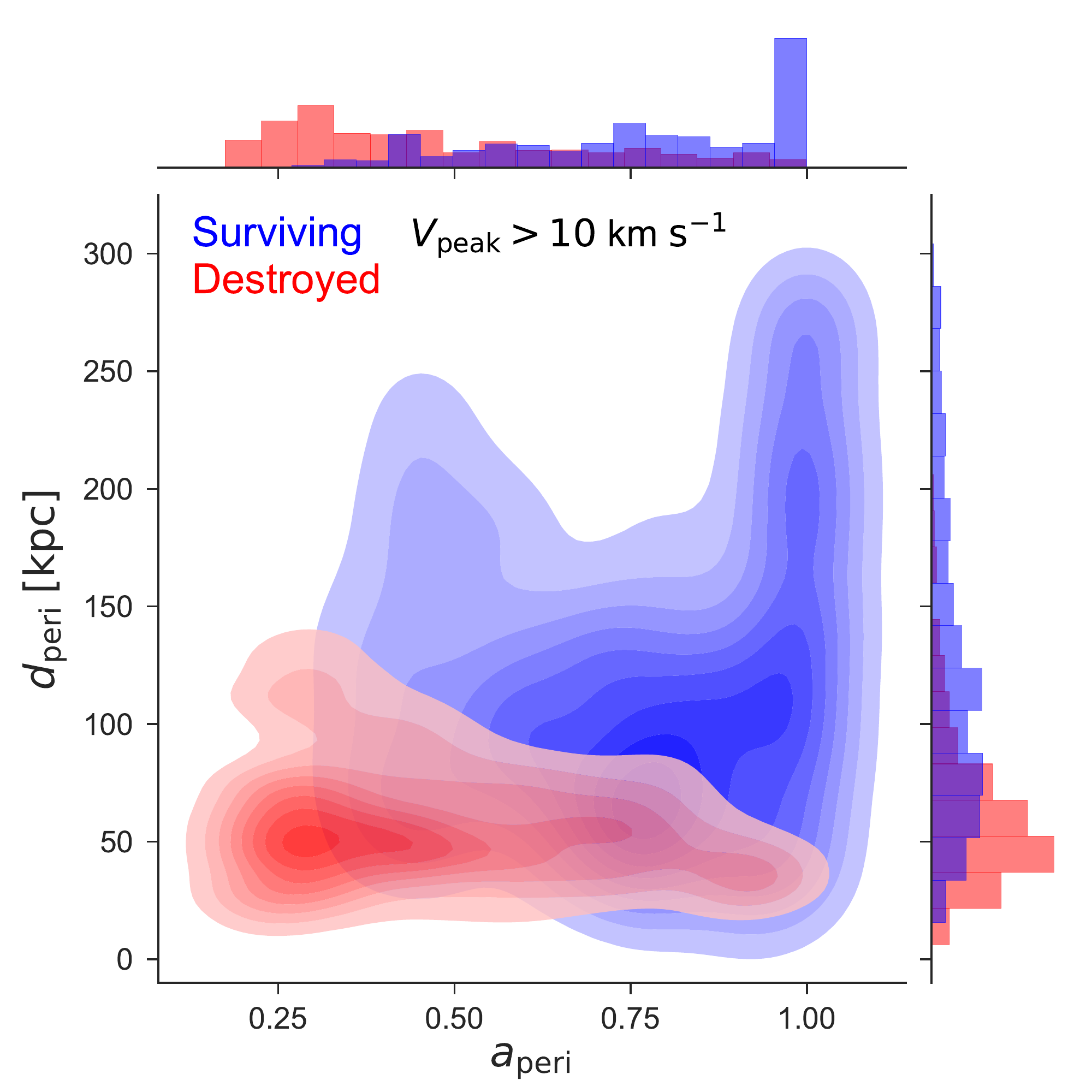}
\caption{Normalized joint and marginal distributions of pericentric distance and scale factor at first pericentric passage after accretion for surviving (blue) and disrupted (red) subhalos with $V_{\rm peak}>10\ \rm{km\ s}^{-1}$ in the \texttt{m12i} and \texttt{m12f} FIRE simulations. We select disrupted subhalos starting at $a=0.25$ ($z=3$).}
\label{fig:contour}
\end{figure}

We choose the subhalo properties listed above because we expect them to correlate with subhalo disruption. Several of these properties are motivated by the results in \citetalias{Garrison-Kimmel170103792}, which show that most of the subhalo disruption in \texttt{m12i} and \texttt{m12f} is caused by the central galactic disk in each simulation. For example, Figure \ref{fig:contour} shows the joint and marginal distributions of $d_{\rm peri}$ and $a_{\rm peri}$ for disrupted and surviving subhalos with $V_{\rm peak}>10\ \rm{km\ s}^{-1}$ in \texttt{m12i} and \texttt{m12f}. Disrupted subhalos, shown in red, tend to have closer pericentric passages that occur at earlier times --- or smaller values of $a_{\rm peri}$ --- than their surviving counterparts, which are shown in blue. The $d_{\rm peri}$ distributions make sense physically; subhalos that pass close to the center of the host experience significant tidal forces due to the galactic disk and are therefore more likely to disrupt.\footnote{\citetalias{Garrison-Kimmel170103792} found that the amount of disruption is largely insensitive to the shape and mass of the central disk, so subhalo disruption in these simulations is at least partly due to disk shocking rather than tidal stripping.} Next, consider the $a_{\rm peri}$ dependence: subhalos that reach pericenter earlier have relatively low masses at pericenter and tend to experience more pericentric passages, both of which contribute to enhanced disruption. Although $a_{\rm peri}$ and $a_{\rm acc}$ are somewhat degenerate properties, we find that including $a_{\rm acc}$ improves our results, likely because subhalos accreted at higher redshifts are tidally stripped for longer periods of time, making them more susceptible to disruption.

\begin{figure}[t]
\centering
\includegraphics[scale=0.43]{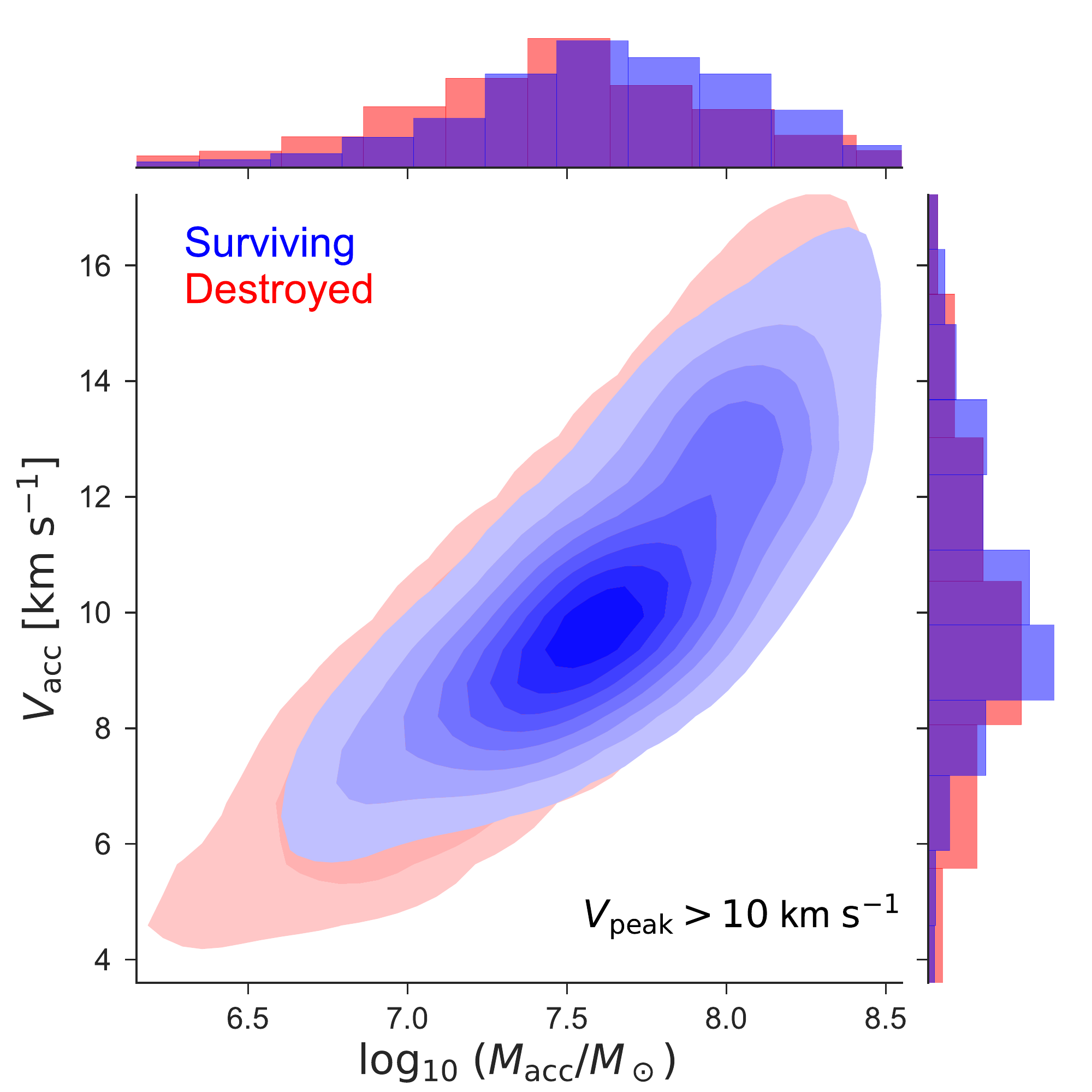}
\caption{Normalized joint and marginal distributions of maximum circular velocity and virial mass at accretion for surviving (blue) and disrupted (red) subhalos with $V_{\rm peak}>10\ \rm{km\ s}^{-1}$ in the \texttt{m12i} and \texttt{m12f} FIRE simulations. We select disrupted subhalos starting at $a=0.25$ ($z=3$); note that $V_{\text{acc}}<V_{\text{peak}}$ for subhalos that are stripped prior to infall (e.g., see \citealp{Behroozi13102239}).}
\label{fig:contourV}
\end{figure}

\begin{center}
\begin{table*}
\centering
\vspace{1.5mm}
\begin{tabular}{ |l|c|c|c| }
\hline
Training Features & OOB Score & Classification Accuracy (Disrupted) & Classification Accuracy (Surviving) \\ \hline
$d_{\rm peri}$ 
& $72\%$ 
& $80\%\pm 3\%$
& $58\%\pm 4\%$\\
\hline
$d_{\rm peri},\ a_{\rm peri}$
& $82\%$ 
& $88\%\pm 2\%$
& $72\%\pm 3\%$\\
\hline
$d_{\rm peri},\ a_{\rm peri},\ a_{\rm acc}$
& $85\%$ 
& $87\%\pm 2\%$
& $82\%\pm 4\%$\\
\hline
$d_{\rm peri},\ a_{\rm peri},\ a_{\rm acc},\ M_{\rm acc}$
& $85\%$ 
& $88\%\pm 2\%$
& $81\%\pm 3\%$\\
\hline
$d_{\rm peri},\ a_{\rm peri},\ a_{\rm acc},\ M_{\rm acc},\ V_{\rm acc}$
& $85\%$ 
& $89\%\pm 2\%$
& $80\%\pm 4\%$\\
\hline
\end{tabular}\vspace{1.5mm}
\caption{Performance metrics for five different random forest classifiers trained on disrupted and surviving subhalos from the \texttt{m12i} and \texttt{m12f} FIRE simulations with $V_{\rm peak}>10\ \rm{km\ s}^{-1}$. The first column lists the subhalo features used to train each classifier. The second column lists the out-of-bag classification score, which is the percentage of subhalos in the training data identified correctly when each tree does not vote on subhalos in its own training set. The third and fourth columns list the percentage of disrupted and surviving subhalos in the test set that are identified correctly by each classifier, averaged over $100$ test-training splits. The test set is the collection of subhalos from the \texttt{m12i} and \texttt{m12f} FIRE simulations with $V_{\rm peak}>10\ \rm{km\ s}^{-1}$ that are not included in the training set. We also indicate the standard deviation of each classification accuracy. Note that the ratio of disrupted to surviving subhalos in our fiducial halo catalog is roughly 3:2.}
\label{tab:percent}
\end{table*}
\end{center}

Figure \ref{fig:contourV} illustrates the $V_{\rm acc}$ and $M_{\rm acc}$ distributions for disrupted and surviving subhalos in \texttt{m12i} and \texttt{m12f}. Interestingly, even though these features mainly contain information about internal rather than orbital subhalo properties, they are useful for identifying disrupted subhalos; as we show below, these properties account for $16\%$ of the total feature importance score for our fiducial five-property classifier. At the low-mass end of the subhalo population, subhalos with lower values of $V_{\rm acc}$ are more likely to be disrupted. In particular, the survival of low-mass subhalos at fixed $M_{\text{acc}}$ is dictated by tidal effects that preferentially disrupt lower-concentration subhalos, i.e., subhalos with smaller values of $V_{\rm acc}$ at fixed $M_{\text{acc}}$. However, at the high-mass end of the subhalo population, subhalos with larger values of $V_{\rm acc}$ are more likely to be disrupted. This behavior suggests that baryonic mechanisms, in addition to the tidal effects of the central disk, contribute to subhalo disruption in the FIRE simulations. Specifically, it is plausible that $V_{\rm acc}$ and $M_{\rm acc}$ encode information about stellar feedback, which can soften central density cusps. In particular, we expect high-mass subhalos with larger values of $V_{\text{acc}}$ to host more massive galaxies and to experience more significant baryonic feedback, i.e., high-mass subhalos with larger values of $V_{\rm acc}$ are more likely to be disrupted. Thus, even though $M_{\rm acc}$ and $V_{\rm acc}$ are highly correlated, it is useful to train on both properties because subhalo concentration determines $V_{\rm acc}$ at fixed $M_{\rm acc}$ and provides physical information about whether a subhalo subject to given tidal forces is disrupted. The advantage of random forest classification is that it captures these complex relationships between subhalo properties and subhalo disruption.

\subsection{Training and Validation}

To train our classifier, we use the \texttt{GridSearchCV} function to search the space of random forest hyperparameters and select the ones that yield the highest out-of-bag (OOB) classification score averaged over ten cross-validation folds of the training data.\footnote{In $n$-fold cross-validation, the training set is divided into $n$ subsets of equal size; $n-1$ of these subsets are used for training, the remaining subset is used for cross-validation, and this procedure is repeated once for each possible cross-validation subset.} These hyperparameters include the number of trees in the forest, the depth of each tree, the maximum number of features used by each tree, and the loss function. We train the classifier using a randomly selected $75\%$ of the disrupted and surviving subhalos from our fiducial training set, with replacement. The number of folds and the ratio of the test-training split do not affect our results. The raw percentage of subhalos with $V_{\rm peak}>10\ \rm{km\ s}^{-1}$ from the hydrodynamic \texttt{m12i} and \texttt{m12f} simulations that are identified correctly by our classifier is $95\%$. We cross-validate this result by computing the OOB classification score, which is defined as the percentage of subhalos from the training data that the random forest classifies correctly when each tree does not vote on subhalos in its own training set. The optimal OOB score for our fiducial five-property classifier is $85\%$, and we find that at least $20$ trees are needed to achieve this OOB score. Our classifier therefore identifies subhalos accurately, although the gap between the overall classification accuracy and the OOB scores suggests that we mildly overfit the training data. In particular, the raw accuracy is higher than the OOB score because decision trees are allowed to vote on subhalos within their respective training sets when classifying all subhalos. To illustrate the relative importance of each subhalo feature, Table \ref{tab:percent} shows the OOB score along with the percentage of correct and incorrect predictions for subhalos in the test set, which is the set of all subhalos that are not included in the training set, for five different classifiers. We calculate these scores for each classifier by using the hyperparameters determined by \texttt{GridSearchCV} and averaging the results over $100$ test-training splits. Each row of Table \ref{tab:percent} lists the results for a classifier trained using an additional subhalo feature; as we add training features, the OOB score and the total classification accuracy generally improve. Note that there are more disrupted subhalos than surviving subhalos in our fiducial training set, so the raw classification accuracy for each set of features is higher than the mean classification accuracy inferred from Table \ref{tab:percent}. Thus, while the classification accuracy for surviving subhalos decreases when $M_{\rm acc}$ and $V_{\rm acc}$ are added, the increase in classification accuracy for disrupted subhalos outweighs this effect. We emphasize, however, that $d_{\text{peri}}$, $a_{\text{peri}}$, and $a_{\text{acc}}$ contain most of the information about subhalo disruption in \texttt{m12i} and \texttt{m12f}.

\begin{figure}
\centering
\includegraphics[scale=0.45]{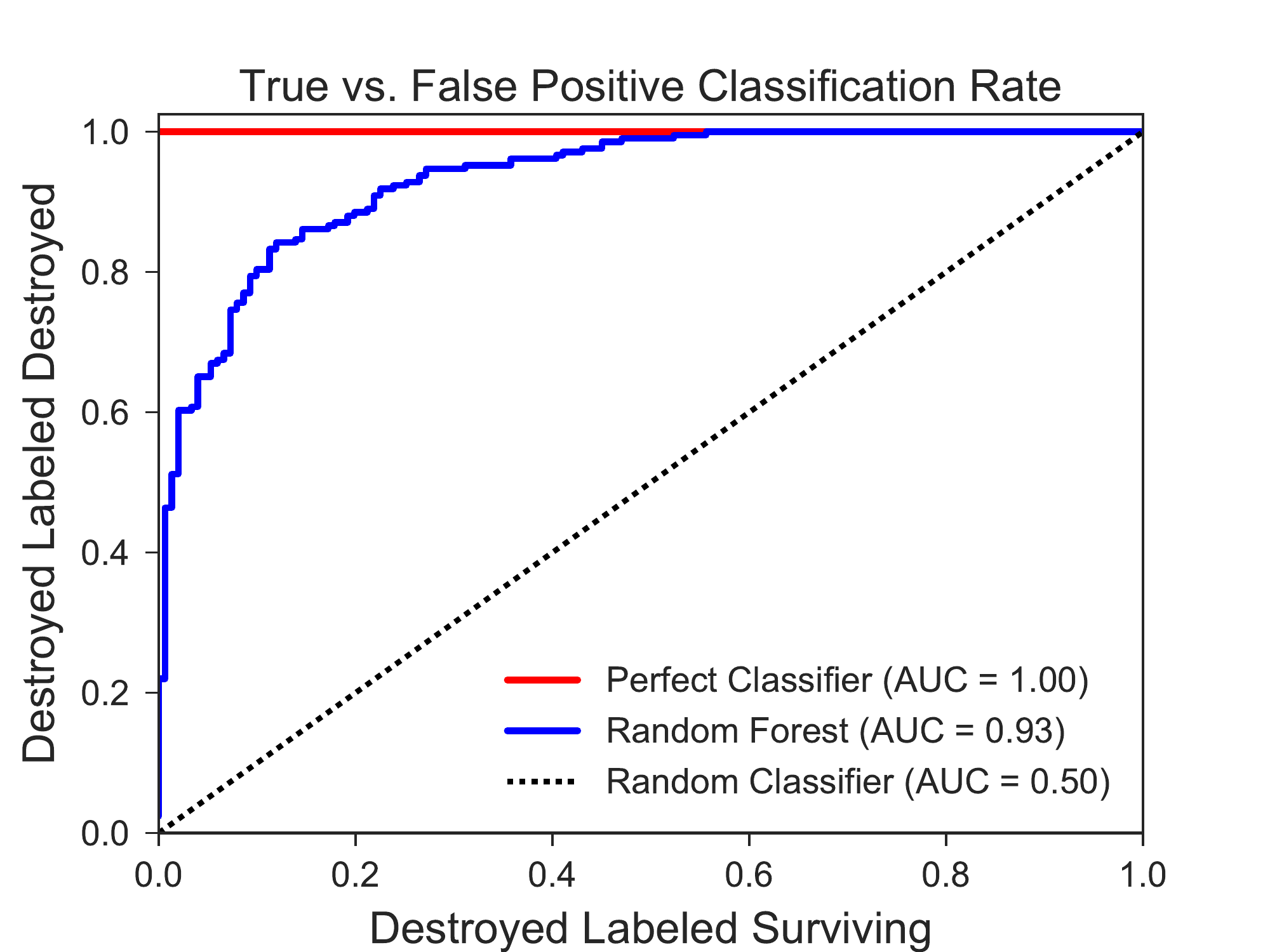}
\caption{True- vs.\ false-positive classification rate for our fiducial five-property random forest classifier, which is trained on subhalos from the \texttt{m12i} and \texttt{m12f} FIRE simulations with $V_{\rm peak}>10\ \rm{km\ s}^{-1}$. These classification rates apply to subhalos that are not included in the training set. The AUC is equal to $1$ for a perfect classifier (red), $0.93$ for our random forest classifier (blue), and $0.5$ for a random classifier (black).}
\label{fig:roc}
\end{figure}

\begin{figure}
\centering
\includegraphics[scale=0.45]{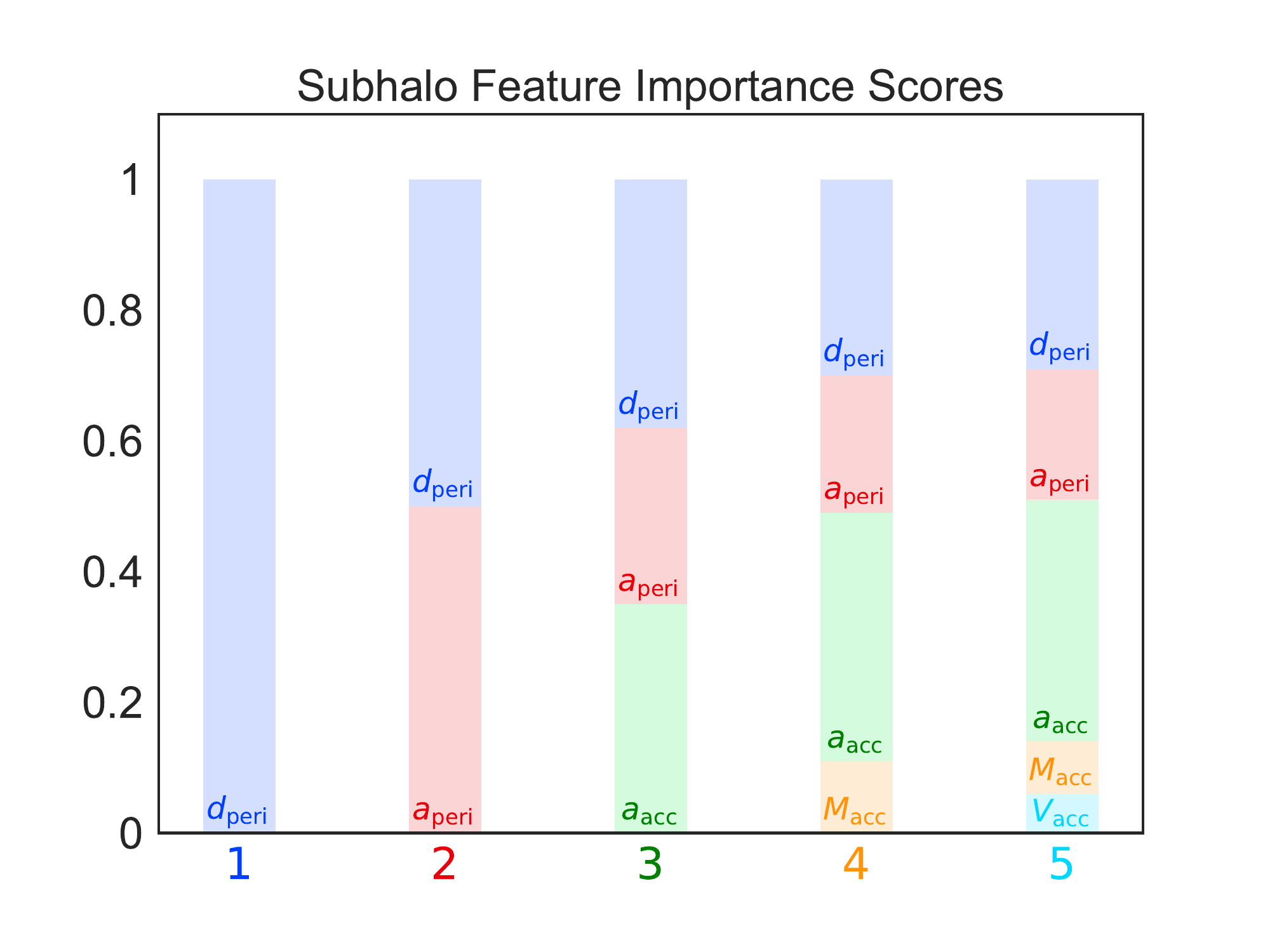}
\caption{Feature importance scores for the five subhalo properties used to classify disrupted and surviving subhalos in the \texttt{m12i} and \texttt{m12f} FIRE simulations. The colored bars above each property indicate the feature importance scores averaged over $100$ test-training splits when that property is added to the training features. Thus, the columns correspond to the five different classifiers in Table~\ref{tab:percent}. For a given classifier, each property's score indicates its relative importance for classifying disrupted and surviving subhalos. Here $d_{\rm peri}$ and $a_{\rm peri}$ are the pericentric distance and scale factor at first pericentric passage after accretion, and $a_{\rm acc}$, $M_{\rm acc}$, and $V_{\rm acc}$ are the scale factor, virial mass, and maximum circular velocity at accretion.}
\label{fig:barplot}
\end{figure}

Next, we examine our classifier's receiver operating characteristic (ROC) curve, which illustrates the rate of true- versus false-positive classifications for subhalos in the test set. The ROC Curve for our five-property classifier is shown in Figure \ref{fig:roc}. The red and black lines illustrate perfect (100$\%$ true-positive rate) and random (true-positive rate equal to false-positive rate) classifiers. We quantify our classifier's performance by calculating the area under the ROC curve (AUC), which confirms that the random forest classifies subhalos in the FIRE simulations accurately: its AUC is $0.93$, while a random classifier has an AUC equal to $0.5$ and a perfect classifier has an AUC equal to $1$. Note that Figure~\ref{fig:roc} shows the ROC curve for a particular test-training split, but the scatter in the ROC curves for different test-training splits is small.

\subsection{Importance of Subhalo Features}

The random forest algorithm determines the feature importance of the various subhalo properties included in the training process. The feature importance indicates the relative importance of each property for predicting whether a given subhalo is disrupted or whether it survives until $z=0$. In particular, a property's feature importance score is the suitably normalized change in the OOB classification score when the property is randomly shuffled among the subhalos in the training set. Thus, the property with the highest feature importance score is the most important for classifying disrupted and surviving subhalos in the \texttt{m12i} and \texttt{m12f} simulations. Figure \ref{fig:barplot} illustrates the mean feature importance scores for each classifier listed in Table \ref{tab:percent}; for a given classifier, the scores are averaged over $100$ test-training splits, and the same hyperparameters are used for each realization. For our fiducial five-property classifier, which corresponds to the fifth column of Figure \ref{fig:barplot}, we find mean feature importance scores of $0.28$ for $d_{\rm peri}$, $0.21$ for $a_{\rm peri}$, $0.35$ for $a_{\rm acc}$, $0.08$ for $M_{\rm acc}$, and $0.08$ for $V_{\rm acc}$. The variance in the feature importance scores for different test-training splits is small, and the scores depend very weakly on the random forest hyperparameters.

Figure \ref{fig:barplot} shows that pericentric distance is an important property for determining whether a given subhalo is disrupted; subhalos with close pericentric passages are more likely to be destroyed. The scale factors at accretion and at first pericentric passage after accretion are also important features. In particular, subhalos that accrete and reach pericenter earlier are preferentially disrupted. The fact that $a_{\text{acc}}$ has the highest feature importance score suggests that the number of pericentric passages, rather than the distance and scale factor associated with each individual passage, is most strongly correlated with subhalo disruption. However, we note that interpreting the feature importance scores for $d_{\rm peri}$ and $a_{\rm peri}$ is complicated by the fact that we defined these properties as the instantaneous distance and scale factor at the final available snapshot for subhalos on infalling orbits that have not reached their true pericenter. The true pericenters for such subhalos occur at smaller values of $d_{\rm peri}$ and larger values of $a_{\rm peri}$ than we have assigned here; in a more detailed analysis, we would need to calculate these features by fitting individual subhalo orbits. However, the fraction of disrupted (surviving) subhalos in our fiducial training set that have not reached their true pericenter by the time of disruption ($z=0$) is only $17\%$ ($20\%$), so the feature importance for $d_{\rm peri}$ and $a_{\rm peri}$ is reasonably accurate.

\subsection{Model Limitations}

Finally, we note that our classification method, like any other model, has its limitations. In particular,
\begin{enumerate}
\item our classifier is only trained on two zoom-in simulations of MW-mass host halos with a specific baryonic physics prescription, and thus it is not clear how well our algorithm will perform on subhalo populations associated with higher- or lower-mass host halos;
\item neither of the hosts that we train on experience a recent major merger, so our classifier might not apply to halos with significantly different formation histories;
\item both hosts form a central galactic disk that is responsible for most of the subhalo disruption, so our classifier mainly captures the dynamical effects of a central disk.
\end{enumerate}
We discuss these limitations in more detail and comment on how they might affect our results in the following section.


\section{Results}
\label{Results}

We now present our main results. In Section \ref{FIREML}, we use our classifier to identify subhalos in DMO simulations of \texttt{m12i} and \texttt{m12f} that are likely to be disrupted in hydrodynamic resimulations. We analyze our predicted surviving subhalo populations by comparing the velocity functions and radial distributions to those from the FIRE, DISK, and DMO simulations in \citetalias{Garrison-Kimmel170103792}. In Section \ref{MWZoomin}, we predict surviving subhalo populations for the suite of DMO zoom-in simulations of MW-mass host halos from Mao et al.\ (\citeyear{Mao150302637}), and we study the resulting velocity functions, radial distributions, and implications for satellite searches.

\subsection{Predictions for DMO Simulations of the FIRE Halos}
\label{FIREML}

\subsubsection{Subhalo Feature Distributions}

There are about twice as many surviving subhalos at $z=0$ in the DMO simulations of \texttt{m12i} and \texttt{m12f} as in the corresponding hydrodynamic simulations. As we have discussed, we expect many of these subhalos to be disrupted due to baryonic effects, including stellar feedback, enhanced tidal stripping, and disk shocking, and our random forest classifier can identify such subhalos based on their internal and orbital properties. In particular, to identify subhalos in the \texttt{m12i} and \texttt{m12f} DMO simulations that are likely to be disrupted by baryonic effects, we select subhalos with $V_{\rm peak}>10\ \rm{km\ s}^{-1}$ at $z=0$, and we use our trained classifier to predict whether these subhalos should have been destroyed at some earlier time using their values of $d_{\rm peri}$, $a_{\rm peri}$, $a_{\rm acc}$, $M_{\rm acc}$, and $V_{\rm acc}$. Note that this method does not require matching subhalos between DMO and hydrodynamic simulations.

\begin{figure}
\centering
\includegraphics[scale=0.43]{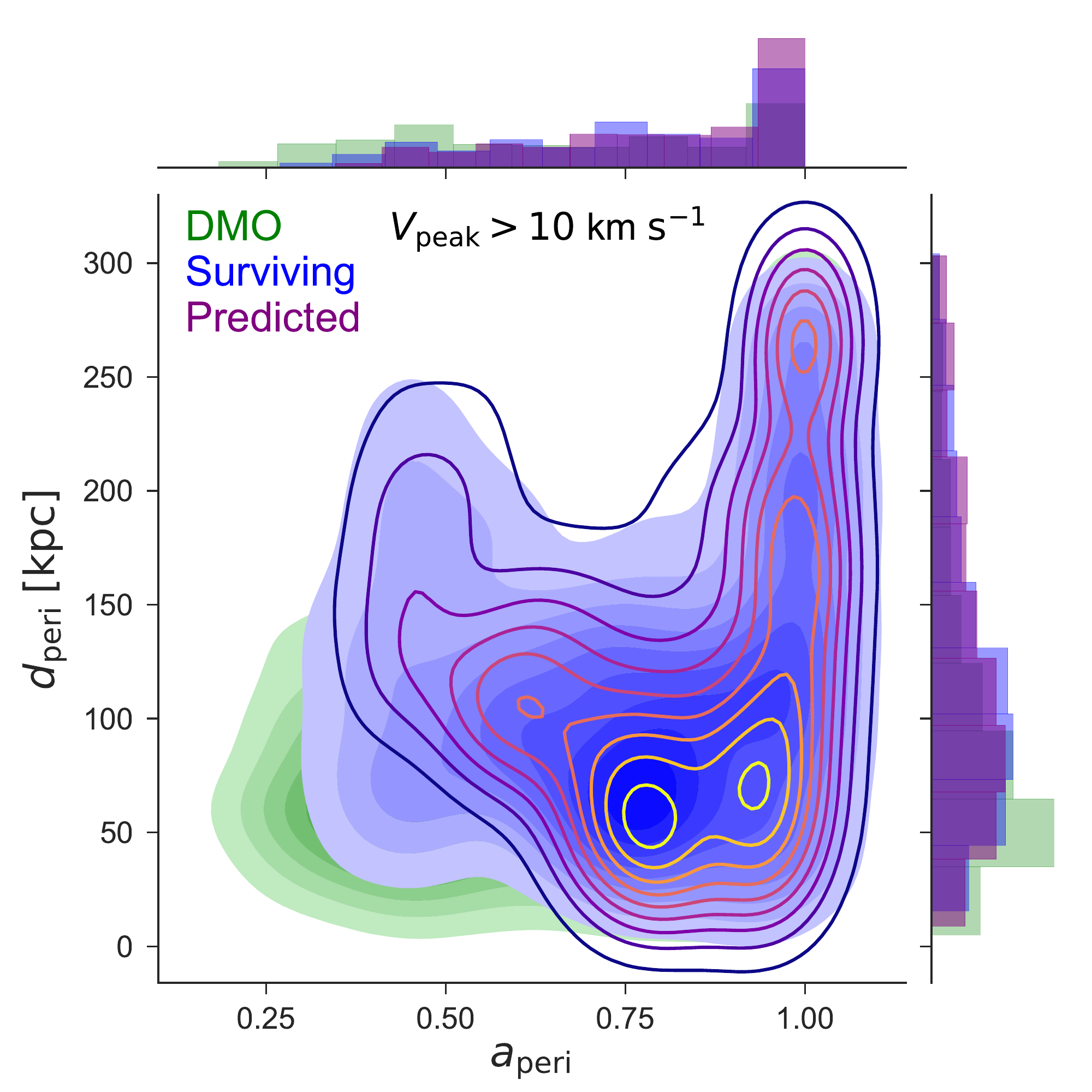}
\caption{Normalized joint and marginal distributions of pericentric distance and scale factor at first pericentric passage after accretion for surviving subhalos in the \texttt{m12i} and \texttt{m12f} FIRE simulations (blue); surviving subhalos from the corresponding DMO simulations are shown in green. The unfilled contour and purple histograms show the most probable surviving subhalo population from the \texttt{m12i} and \texttt{m12f} DMO simulations predicted by our random forest classifier.}
\label{fig:predictions2}
\end{figure}

Figure \ref{fig:predictions2} shows the joint and marginal distributions of $d_{\rm peri}$ and $a_{\rm peri}$ for surviving subhalos from the \texttt{m12i} and \texttt{m12f} DMO simulations predicted by our random forest classifier. The random forest predicts a surviving subhalo population in $d_{\rm peri}-a_{\rm peri}$ space that agrees well with the hydrodynamic data; we also find good agreement in the spaces defined by the other subhalo features. Of course, since our classifier is trained on subhalos from the \texttt{m12i} and \texttt{m12f} FIRE simulations, we expect it to perform particularly well on the corresponding DMO simulations, which have identical initial conditions. Nevertheless, these results are encouraging: even though there is no galactic disk or stellar feedback in the DMO simulations, our classifier efficiently predicts subhalo populations that are in good agreement with the hydrodynamic results. In particular, once the classifier has been trained on the hydrodynamic simulations, it can immediately predict surviving subhalo populations from DMO halo catalogs. Simulations that include baryonic effects by hand, such as the DISK simulations presented in \citetalias{Garrison-Kimmel170103792}, are complementary to our approach, since they provide more direct physical modeling at the expense of increased computational costs.

In general, at least three mechanisms contribute to enhanced subhalo disruption in the \texttt{m12i} and \texttt{m12f} hydrodynamic simulations relative to the DMO simulations: tidal effects due to the central galactic disk, stellar feedback, and characteristic changes in subhalo orbits due to the presence of baryons. The results from the DISK simulations in~\citetalias{Garrison-Kimmel170103792} indicate that the central disk is the main source of subhalo disruption in these simulations, but the frequency of disruption events might be enhanced by stellar feedback, which can soften central density cusps \citep{Governato12020554,Pontzen11060499,Zolotov12070007,DiCintio14045959}; as noted above, we multiply all circular velocities in the DMO and DISK simulations by a factor of $\sqrt{1-f_{\rm b}}$ to approximate this effect. Meanwhile, Zhu et al.\ (\citeyear{Zhu170105933}) analyzed the orbital properties of subhalos in hydrodynamic and DMO zoom-in simulations of an MW-mass host halo from the Aquarius Project and found that the distributions of subhalos in different orbital families change when baryons are included. It is difficult to assess the importance of the characteristic differences in internal and orbital subhalo properties between hydrodynamic and DMO simulations in general; however, the fact that we predict subhalo feature distributions starting from DMO halo catalogs that agree with hydrodynamic results suggests that these effects are relatively unimportant.

\subsubsection{Subhalo Counts}

\begin{figure*}
\centering
\includegraphics[scale=0.46]{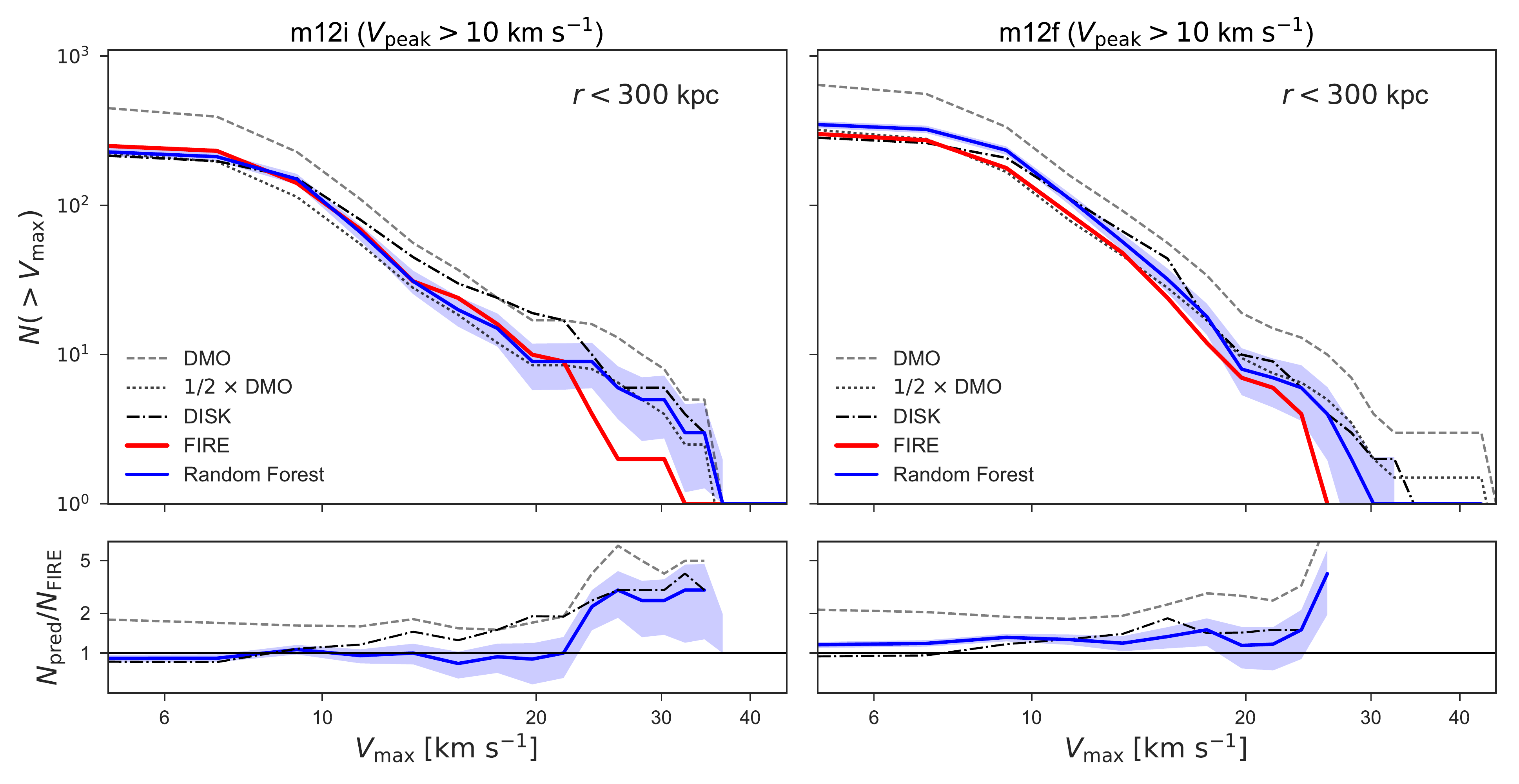}
\includegraphics[scale=0.46]{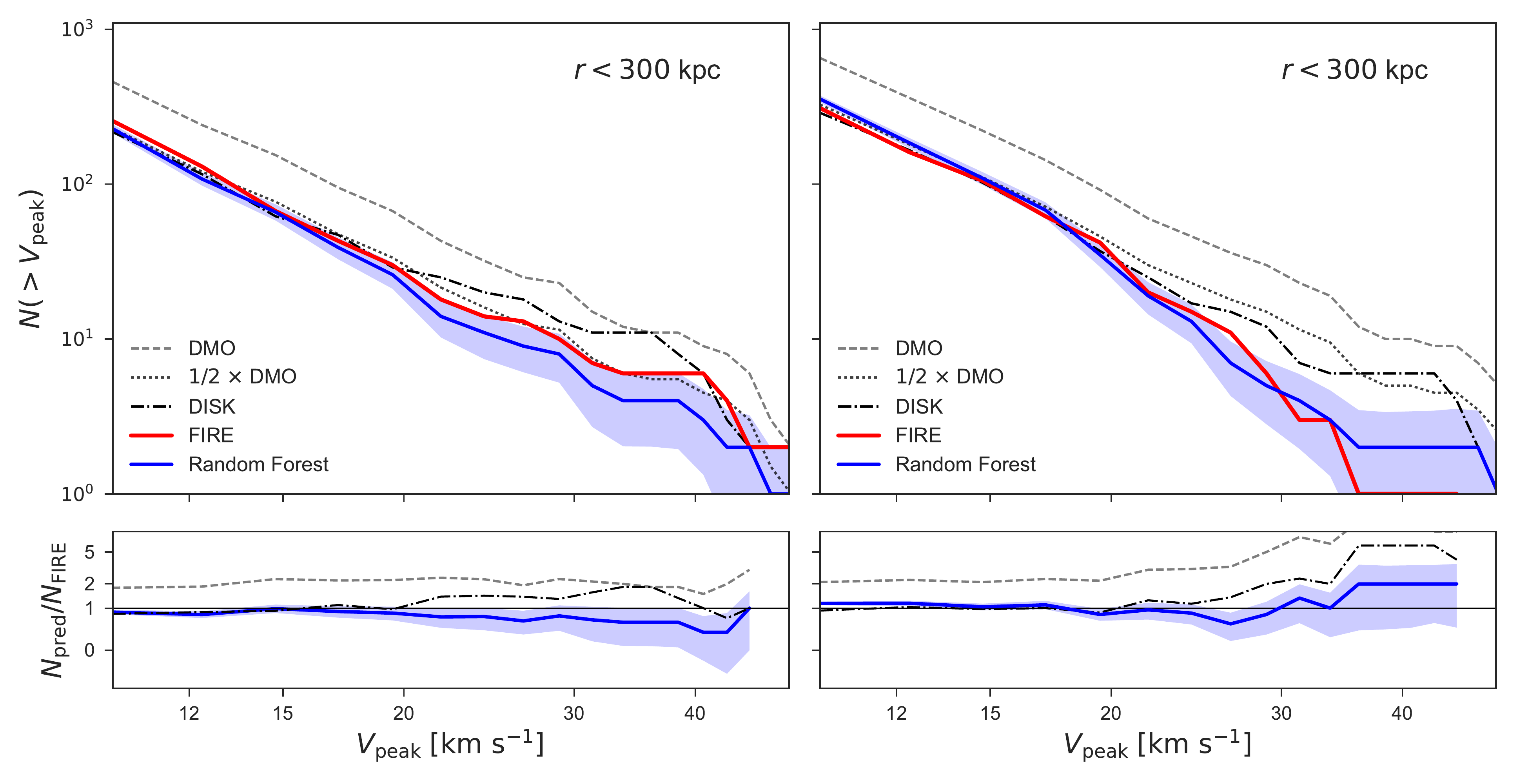}
\caption{Velocity functions for subhalos hosted by \texttt{m12i} (left) and \texttt{m12f} (right), predicted from DMO simulations of these hosts by our random forest classifier (blue). The top panels show velocity functions evaluated using the maximum circular velocity at $z=0$, and the bottom panels show velocity functions evaluated using the peak circular velocity $V_{\rm peak}$. Our classifier is trained on subhalos with $V_{\rm peak}>10\ \rm{km\ s}^{-1}$ from both hydrodynamic simulations. The FIRE (red), DISK (dot-dashed), and DMO (dashed) results are shown for comparison; recall that we scale circular velocities in the DISK and DMO halo catalogs by a factor of $\sqrt{1-f_{\rm b}}$. Dotted lines show the DMO results scaled by a factor of $1/2$ for comparison. We restrict these velocity functions to subhalos within $300\ \rm{kpc}$ of their respective host at $z=0$. The bottom panels show the ratio $N_{\rm pred}/N_{\rm FIRE}$, where $N_{\rm pred}$ is the number of surviving subhalos predicted by the random forest and $N_{\rm FIRE}$ is the number of subhalos in each FIRE simulation. Shaded areas show the standard deviation about the most probable random forest prediction for $1000$ draws from a Poisson distribution with a mean value of $N_{\rm pred}$ at each value of $V_{\rm max}$ or $V_{\rm peak}$.}
\label{fig:vmax}
\end{figure*}

\begin{figure*}
\centering
\includegraphics[scale=0.46]{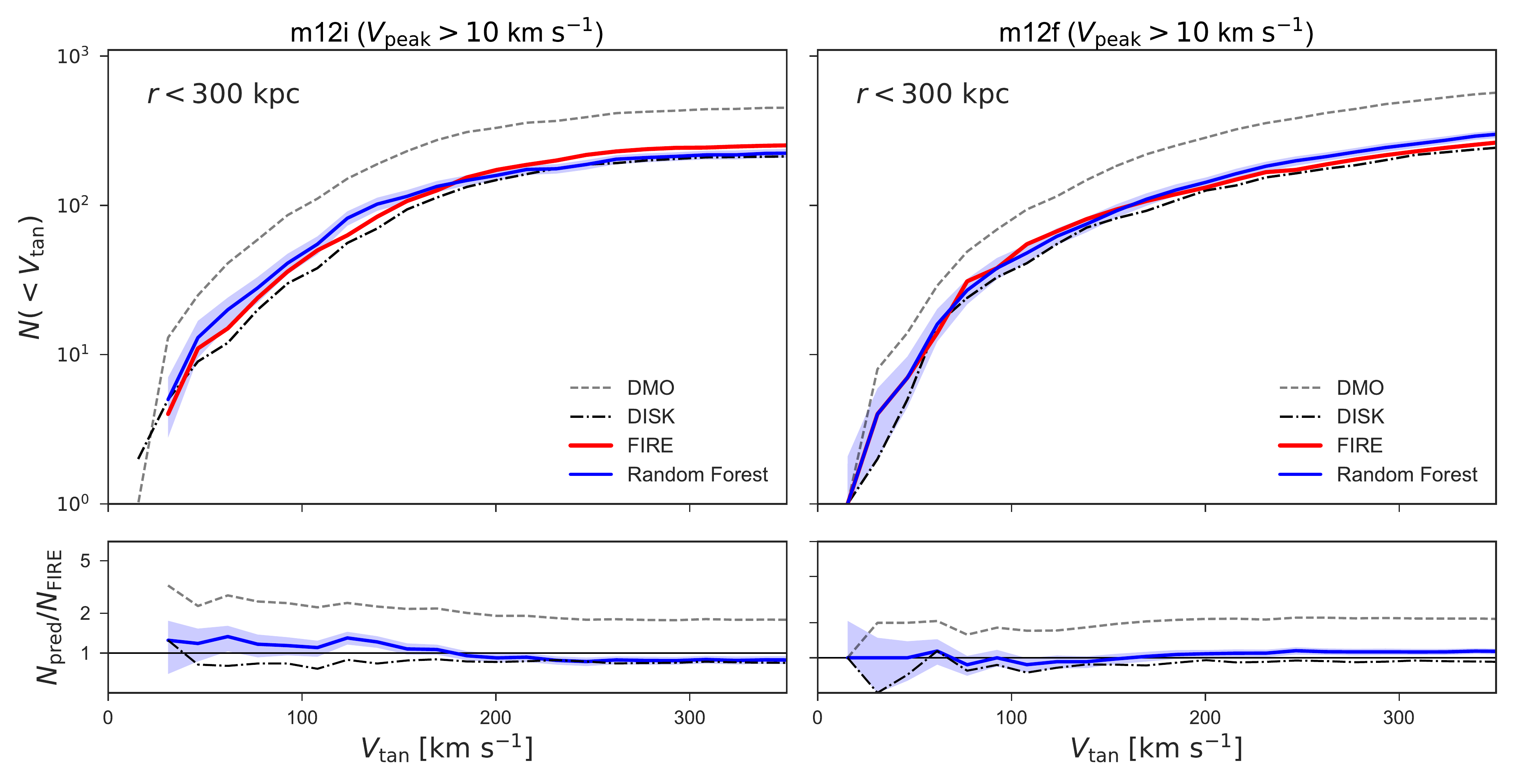}
\includegraphics[scale=0.46]{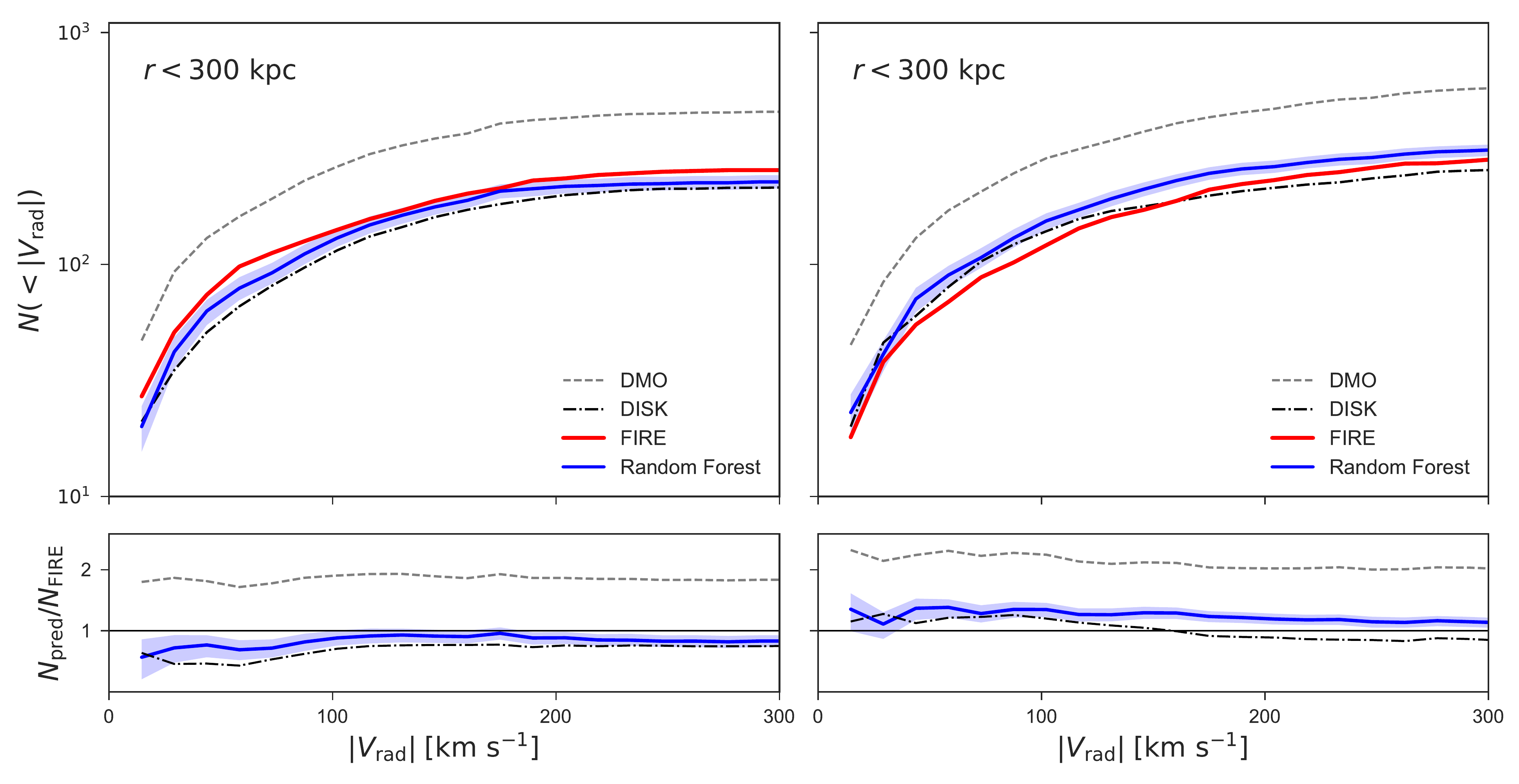}
\caption{Distributions of tangential orbital velocities (top) and radial orbital velocities (bottom) for subhalos with $V_{\rm peak}>10\ \rm{km\ s}^{-1}$ hosted by \texttt{m12i} (left) and \texttt{m12f} (right) at $z=0$, predicted from DMO simulations of these host by our random forest classifier (blue). The classifier is trained on subhalos with $V_{\rm peak}>10\ \rm{km\ s}^{-1}$ from both FIRE simulations. We restrict these distributions to subhalos within $300\ \rm{kpc}$ of their respective hosts at $z=0$. The various curves and panels are described in Figure \ref{fig:vmax}.}
\label{fig:vtan}
\end{figure*}

\begin{figure*}
\centering
\includegraphics[scale=0.46]{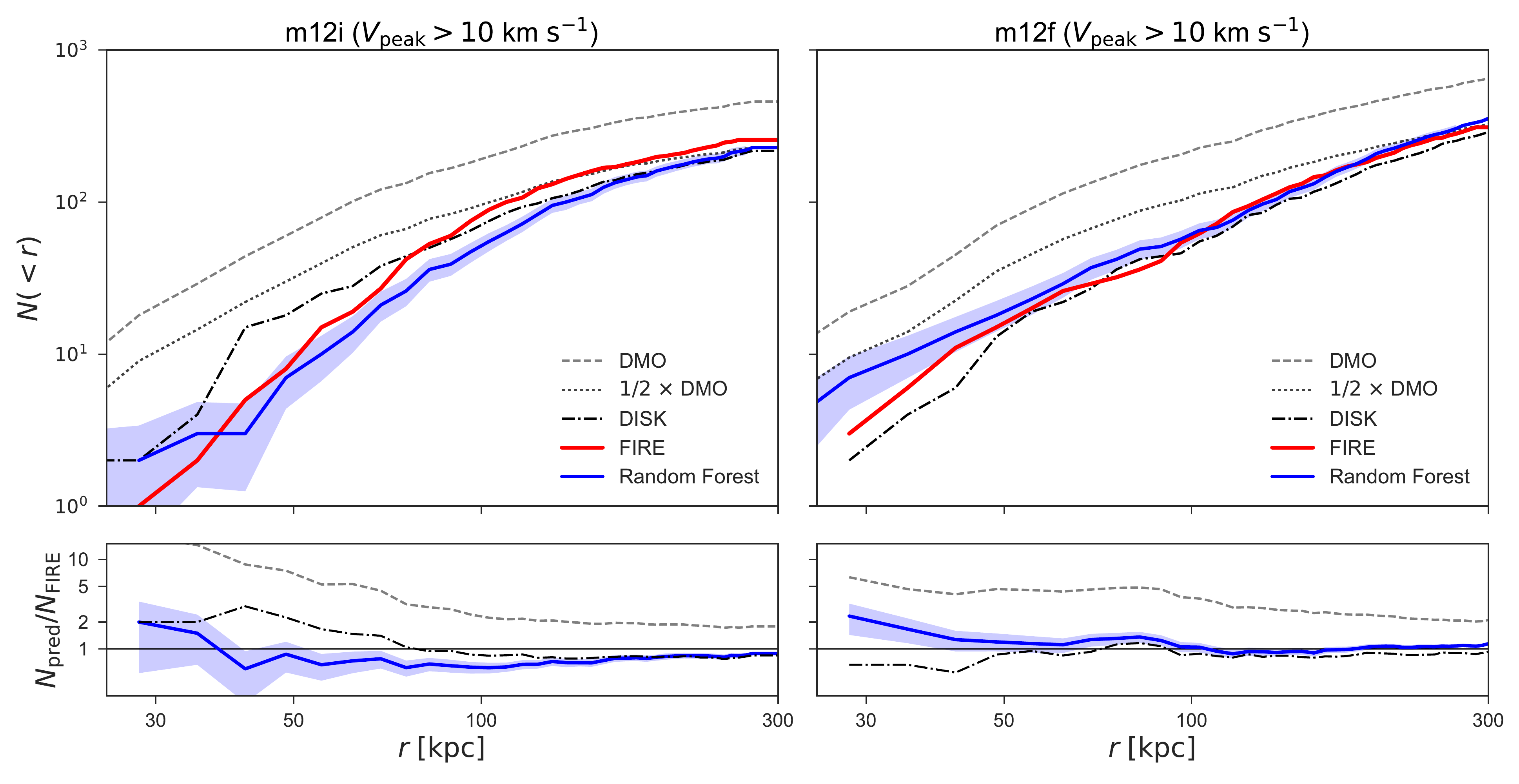}
\caption{Radial distributions of subhalos with $V_{\rm peak}>10\ \rm{km\ s}^{-1}$ hosted by \texttt{m12i} (left) and \texttt{m12f} (right) at $z=0$, predicted from DMO simulations of these hosts by our random forest classifier (blue). The classifier is trained on subhalos with $V_{\rm peak}>10\ \rm{km\ s}^{-1}$ from both FIRE simulations, and the various curves and panels are described in Figure \ref{fig:vmax}. The scaled DMO curve overpredicts the number of surviving subhalos at small radii by an order of magnitude, highlighting the enhanced subhalo disruption in the inner regions of the hydrodynamic simulation due to the central disk.}
\label{fig:radial}
\end{figure*}

Having shown that we can predict the feature distributions of surviving subhalos from DMO simulations of \texttt{m12i} and \texttt{m12f}, we turn to our predictions for the number of surviving subhalos as a function of various properties. In Figure \ref{fig:vmax}, we present our predictions for the \texttt{m12i} and \texttt{m12f} velocity functions; the top panels show the velocity functions evaluated using $V_{\rm{max}}$, and the bottom panels show the velocity functions evaluated using $V_{\rm peak}$. The blue lines show the most probable surviving subhalo populations predicted by our random forest algorithm for each host halo; we also plot the FIRE, DISK, and DMO results for comparison. We restrict the velocity functions to subhalos within $300$ kpc of the center of their respective host at $z=0$, since this roughly corresponds to the virial radii of \texttt{m12i} and \texttt{m12f}. Similarly, Figure~\ref{fig:vtan} shows the distribution of tangential and radial orbital velocities for subhalos within $300\ \rm{kpc}$ of their respective host at $z=0$, and Figure \ref{fig:radial} shows the radial distribution of surviving subhalos at $z=0$ within each host halo. In Figures \ref{fig:vmax}--\ref{fig:radial}, we only include subhalos with $V_{\rm peak}>10\ \rm{km\ s}^{-1}$ to match the cut used in our fiducial training set. The bottom panels in these figures show the number of surviving subhalos predicted by the most probable realization of our random forest classifier divided by the number of subhalos found in each hydrodynamic simulation. We also plot the Poisson error associated with the random forest predictions as shaded areas in each figure. In Appendix \ref{appendix}, we show that the intrinsic scatter in the random forest predictions is small.

There are several interesting aspects of Figures \ref{fig:vmax}--\ref{fig:radial} that are worth exploring. Our random forest algorithm predicts that the amount of substructure in each host is significantly reduced relative to the DMO simulations, bringing the velocity functions and radial distributions into good agreement with the FIRE results. The random forest predictions for the velocity functions are comparable to the DISK simulations at low velocities, which indicates that the effects of the disk are largely encoded in the subhalo properties that we use to train our classifier, at least for subhalos with low values of $V_{\rm max}$ or $V_{\rm peak}$. However, the random forest predicts more subhalo disruption than the DISK simulations for $V_{\rm max}\gtrsim 15\ \rm{km\ s}^{-1}$ or $V_{\rm peak}\gtrsim 20\ \rm{km\ s}^{-1}$ and generally matches the FIRE results more closely in these regimes. The minor discrepancies for $V_{\rm max}\gtrsim 15\ \rm{km\ s}^{-1}$ are likely caused by enhanced mass stripping due to baryonic effects, which would shift the predictions toward smaller velocities at high $V_{\rm max}$. 

Our predicted radial distributions are also generally comparable to the DISK simulations; however, for $30\ \rm{kpc}\lesssim$ $r\lesssim 100\ \rm{kpc}$, where the disk should be particularly effective at disrupting subhalos, our classifier predicts more subhalo disruption than the \texttt{m12i} DISK simulation and matches the FIRE results more closely for both hosts. Finally, Figure \ref{fig:vtan} shows that our classifier predicts a substantial reduction in the number of subhalos with low tangential velocities, even though it is not explicitly trained on orbital velocities. Our predicted tangential and radial velocity distributions are similar to the DISK results for \texttt{m12i}, while we slightly overpredict the number of high-$V_{\rm tan}$ and high-$V_{\rm rad}$ subhalos for \texttt{m12f}. Comparing our predictions to the DISK simulations is a particularly useful way to assess whether our classifier captures baryonic physics beyond the dynamical effects of a central galactic disk, since the DISK simulations do not modify internal subhalo properties. Thus, Figures \ref{fig:vmax}--\ref{fig:radial} suggest that our classifier captures both the tidal effects of a disk and additional baryonic processes that contribute to subhalo disruption.

Our random forest classifier predicts that many subhalos with large values of $V_{\rm max}$ and $V_{\rm peak}$ should be disrupted, while these subhalos are not necessarily destroyed in the DISK simulations (see Figures \ref{fig:vmax} and \ref{fig:comparison}). These subhalos either orbit at large radii, so that they are not significantly affected by the disk, or they are too tightly bound to be disrupted by the disk alone. We find that $45\%$ ($84\%$) of the disrupted subhalos from \texttt{m12i} and \texttt{m12f} with $V_{\rm peak}>20\ \rm{km\ s}^{-1}$ have pericentric passages within $50\ \rm{kpc}$ ($100\ \rm{kpc}$) of their respective hosts. The disk does not seem to be the main factor that contributes to the destruction of the remaining subhalos, though a combination of stellar feedback and tidal forces could lead to their disruption. Interestingly, the region of the $V_{\rm{max}}$ and $V_{\rm{peak}}$ functions where we predict enhanced subhalo disruption relative to the DISK simulations ($V_{\rm max}\gtrsim 15\ \rm{km\ s}^{-1}$ and $V_{\rm peak}\gtrsim 20\ \rm{km\ s}^{-1}$) corresponds to the regime where baryonic physics can efficiently create cored subhalo density profiles \citep{Chan150702282,Tollet150703590,Fitts161102281}. It is also intriguing that our classifier predicts both the $V_{\rm max}$ and $V_{\rm peak}$ functions accurately, even though it does not account for enhanced mass stripping beyond the $\sqrt{1-f_b}$ circular velocity correction, which does not reproduce the hydrodynamic results on its own (for example, compare the `Raw DMO' and `DMO' curves in \citetalias{Garrison-Kimmel170103792}). Since $V_{\rm max}$ and $V_{\rm peak}$ are proxies for satellite luminosity, our method can therefore be extended to predict satellite galaxy populations associated with MW-mass host halos (see Figure~\ref{fig:luminosity}); in addition, it can be used to constrain the cumulative mass functions of dark and luminous substructures relevant to gravitational-lensing analyses. Clearly, a more diverse training sample is required in order to make robust predictions regarding the populations of satellite galaxies around the MW and around the MW analogs from the SAGA survey. Nonetheless, Figures \ref{fig:vmax} and \ref{fig:radial} show that classification algorithms can predict subhalo populations in good agreement with hydrodynamic simulations, providing an efficient way to explore the range of possible satellite galaxy populations associated with a particular host halo.

One could argue that the efficiency of our approach is outweighed by the fact that we must train our classifier on computationally expensive hydrodynamic simulations in order to predict surviving subhalo populations for corresponding DMO simulations. However, as we demonstrate in the following section, our method can be used to predict surviving subhalo populations when hydrodynamic simulations are unavailable. Of course, the surviving subhalo populations we predict in this paper are specific to the FIRE simulations that we use to train our classifier. Nonetheless, even though the generality of our results is limited by the small size of our training set, our work suggests that random forest classification can be used to predict realistic subhalo populations given a sufficiently diverse sample of hydrodynamic training simulations. In addition, we emphasize that our classifier is trained on simulations that yield satellite populations that are consistent with the observed mass functions and velocity dispersion functions for satellites of the MW and M31.

\subsection{Predictions for a Suite of DMO Zoom-in Simulations}
\label{MWZoomin}

\subsubsection{Subhalo Counts}

\begin{figure}
\centering
\includegraphics[scale=0.45]{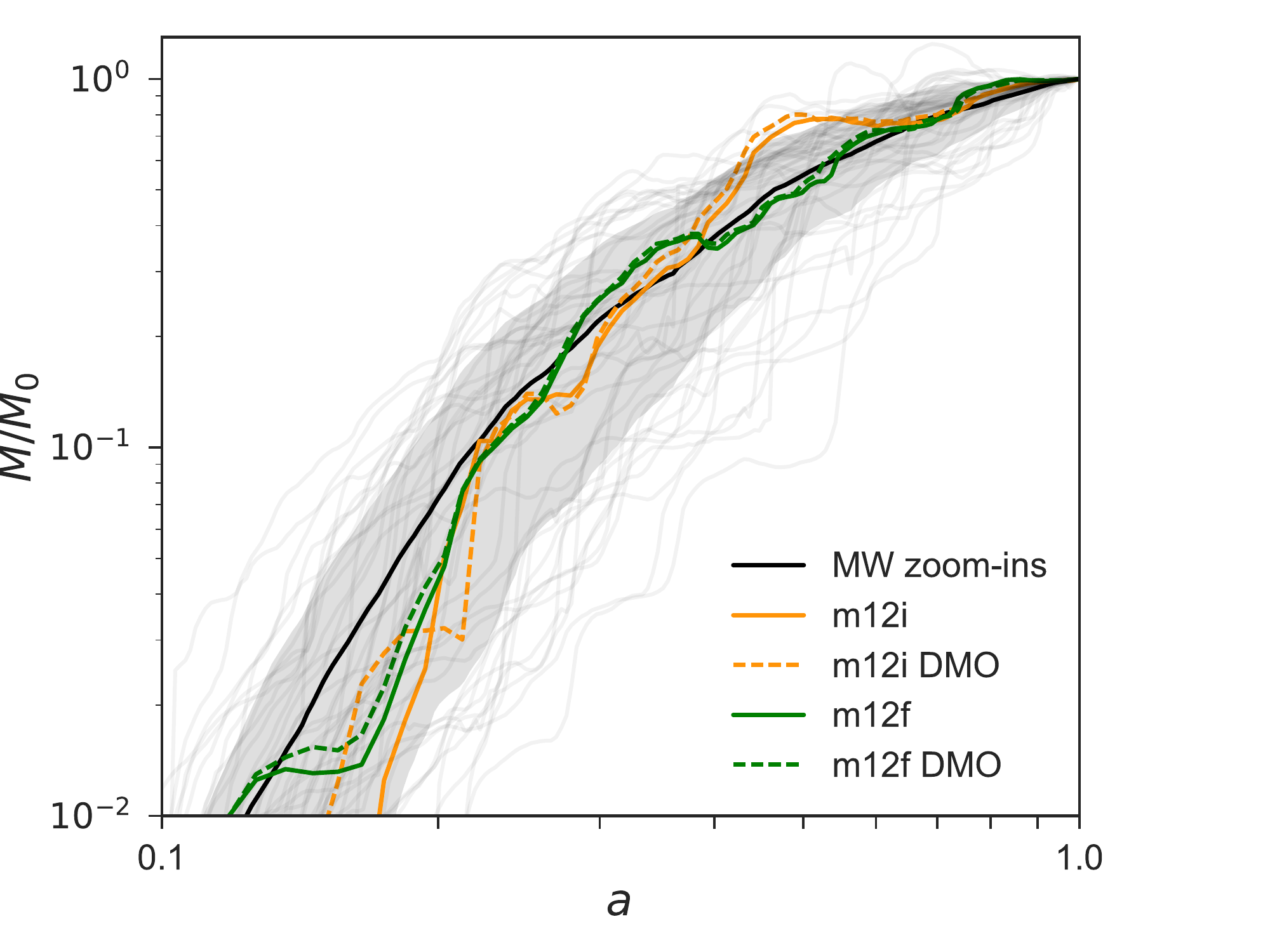}
\caption{Mass accretion histories for the suite of DMO zoom-in simulations of MW-mass host halos presented in Mao et al.\ (\citeyear{Mao150302637}). The black line shows the mean mass accretion history for the $45$ hosts, and the shaded area shows the associated $\pm 1\sigma$ standard deviation. Mass accretion histories for the \texttt{m12i} and \texttt{m12f} FIRE simulations are shown in orange and green, respectively.}
\label{fig:MAH}
\end{figure}

We now use our classifier to identify subhalos from a suite of independent DMO zoom-in simulations that are likely to be disrupted in hydrodynamic resimulations. In particular, we predict surviving subhalo populations for the $45$ zoom-in simulations of MW-mass host halos from Mao et al.\ (\citeyear{Mao150302637}). We refer the reader to Mao et al.\ (\citeyear{Mao150302637}) for a detailed description of the simulations, but we briefly highlight the most important aspects for this work. The host halos lie in the mass range $M_{\rm vir}=10^{12\pm 0.03}\ M_{\rm \odot}$ and have a variety of formation histories; we plot the mass accretion histories for these hosts in Figure \ref{fig:MAH}. Note that \texttt{m12i} and \texttt{m12f} have formation histories that are consistent with these host halos, so we expect our model to perform well on this simulation suite. Of course, our model would not accurately predict subhalo disruption for hosts with significantly different formation histories due to the limited size of our training set. The zoom-in simulations were run at a lower resolution than the DMO simulations of \texttt{m12i} and \texttt{m12f}; the dark matter particle mass is $3\times 10^{5}\ M_{\rm \odot}$, and Mao et al.\ (\citeyear{Mao150302637}) estimated that $V_{\rm max} \sim 9\ \rm{km\ s}^{-1}$ is a conservative lower limit for the subhalo circular velocity resolution. Halo catalogs and merger trees were generated using the \texttt{ROCKSTAR} halo finder and the \texttt{consistent-trees} merger code \citep{Behroozi11104372,Behroozi11104370}. Again, we scale all subhalo masses by $1-f_{\rm b}$ and all subhalo circular velocities by $\sqrt{1-f_{\rm b}}$ in our post-processing of the halo catalogs. As noted at the end of Section \ref{Introduction}, the cosmological parameters for these simulations are slightly different than those used in the FIRE simulations, and we adjust the parameters in our analysis accordingly.

\begin{figure*}
\centering
\includegraphics[scale=0.525]{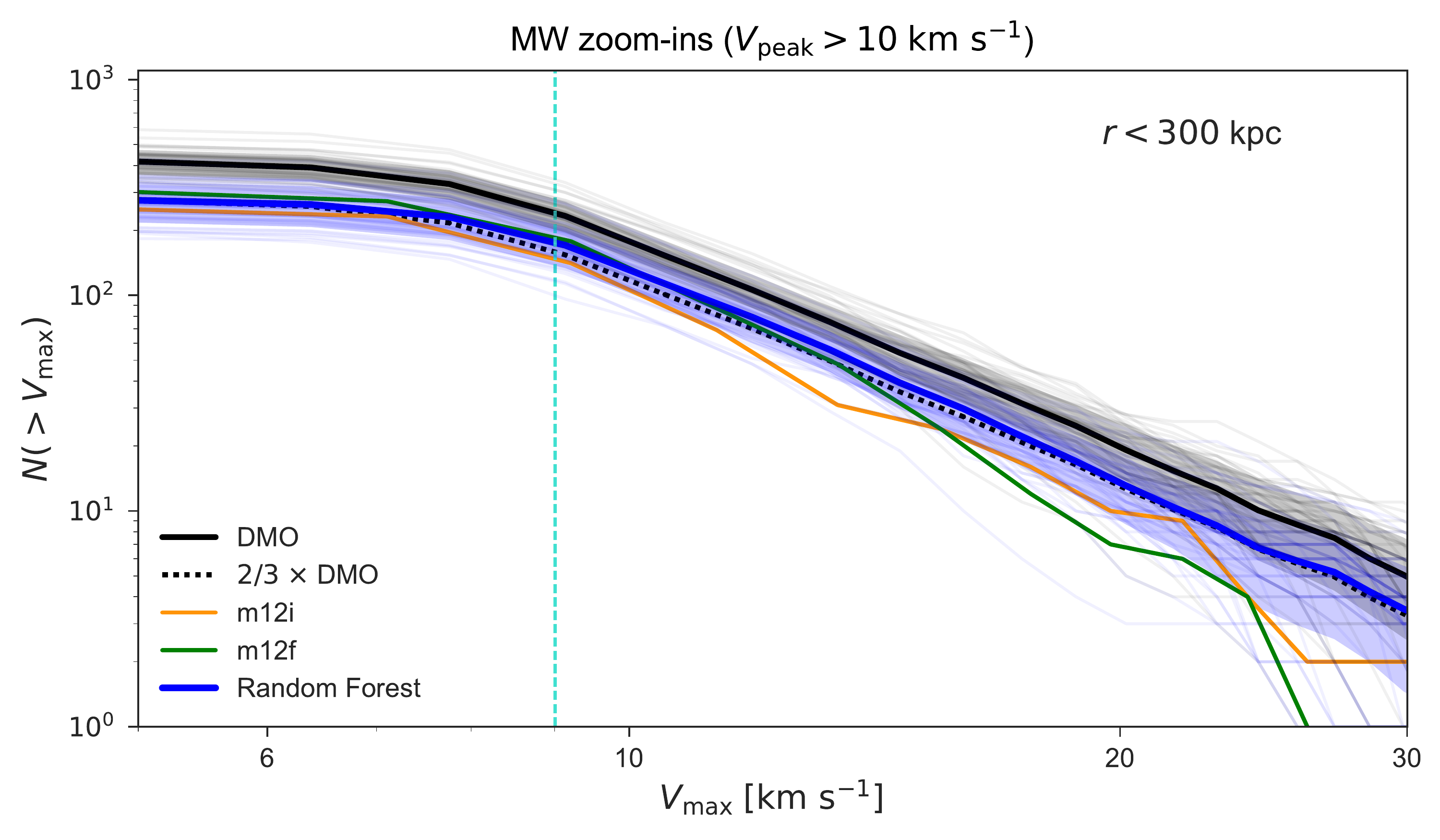}
\\ \ \\
\includegraphics[scale=0.525]{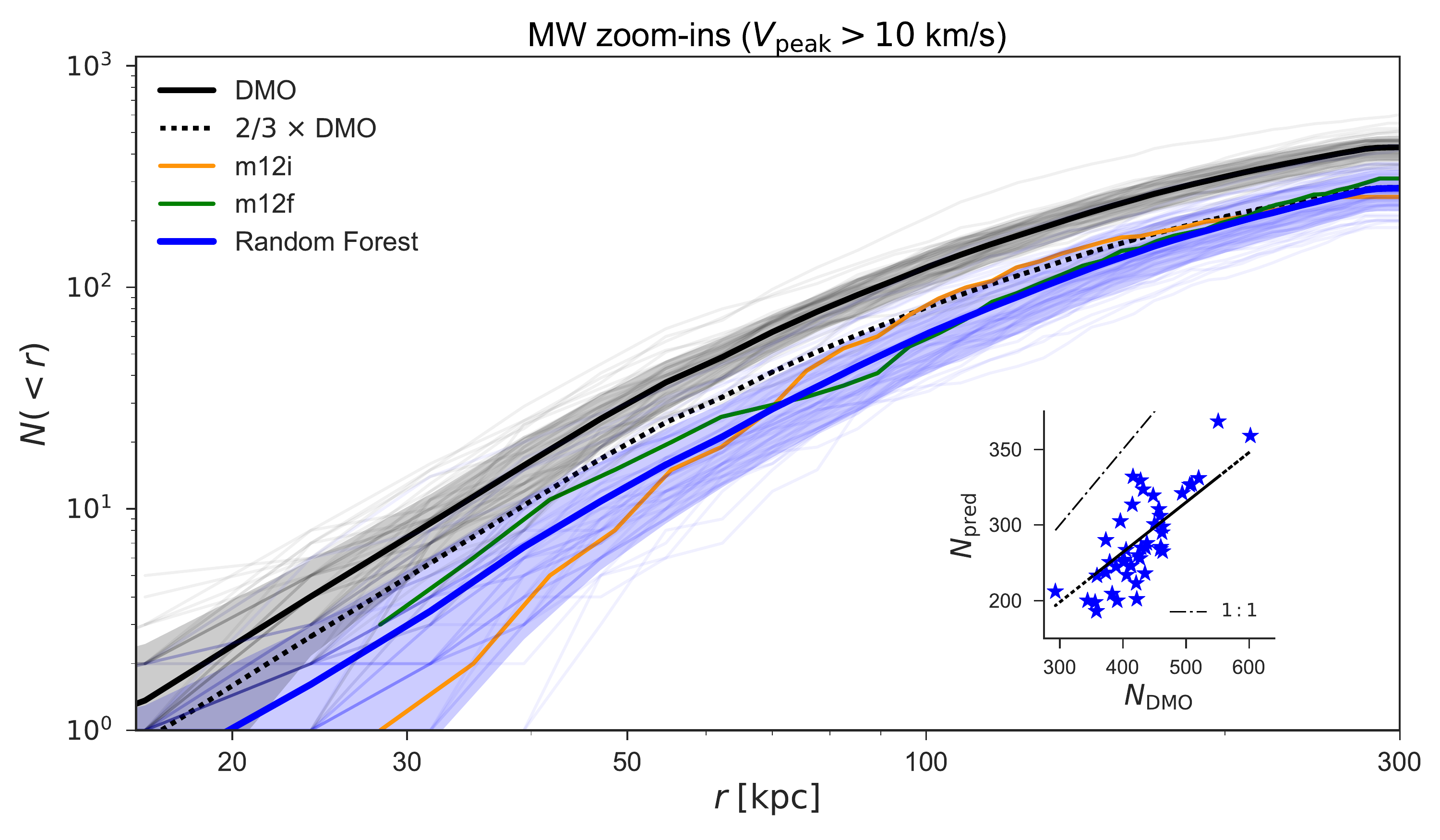}
\caption{Velocity functions (top) and radial distributions (bottom) for the suite of 45 zoom-in simulations of MW-mass host halos presented in Mao et al.\ (\citeyear{Mao150302637}). The thick lines show the mean number of subhalos predicted by the DMO simulations (black) and by our random forest classifier (blue), which is trained on the \texttt{m12i} and \texttt{m12f} FIRE simulations; the shaded areas show the $\pm 1\sigma$ standard deviation of these predictions. The thin lines show the DMO result and the most probable random forest prediction for each host. The thick dotted lines show the mean DMO velocity function and radial distribution scaled by a factor of $2/3$ for visual comparison, and the orange and green lines show the results for \texttt{m12i} and \texttt{m12f}, respectively. Note that the scaled DMO line in the top panel is mostly obscured by the random forest prediction. The inset in the bottom panel shows the number of predicted surviving subhalos with $V_{\rm peak}>10\ \rm{km\ s}^{-1}$ and within $300\ \rm{kpc}$ of their respective host versus the number of such subhalos in the corresponding DMO simulations. The thick dotted line in the inset shows the constant fraction of surviving subhalos corresponding to the scaled DMO curves, and the thin dash-dotted line shows a $1:1$ relationship for comparison. The vertical line at $V_{\rm max}=9\ \rm{km\ s}^{-1}$ in the top panel represents a conservative resolution limit for these simulations.}
\label{fig:zoomins}
\end{figure*}

In Figure \ref{fig:zoomins}, we plot the maximum circular velocity functions and radial distributions for the subhalo populations from this simulation suite, along with those predicted by the most probable realization of our random forest classifier for each simulation. We also plot the results from the \texttt{m12i} and \texttt{m12f} FIRE simulations, along with the mean DMO curves scaled by a constant factor, for comparison. In particular, we scale the mean DMO curves by a factor of $2/3$ so that the average number of subhalos with $V_{\rm peak}>10\ \rm{km\ s}^{-1}$ and $r<300\ \rm{kpc}$ matches the mean random forest prediction. The random forest predictions were generated using the method described above. We classify subhalos in each zoom-in simulation using the features $d_{\rm peri}$, $a_{\rm peri}$, $a_{\rm acc}$, $M_{\rm acc}$, and $V_{\rm acc}$, and we restrict the velocity functions to subhalos within $300\ \rm{kpc}$ of their respective host at $z=0$. We plot the most probable realization of the random forest prediction for each host. The intrinsic scatter in our random forest predictions is small.

Figure \ref{fig:zoomins} shows that the reduction in the total number of subhalos predicted by our random forest classifier is larger than the host-to-host scatter for the subhalo populations in these zoom-in simulations. In particular, our classifier predicts that the total number of subhalos with $V_{\rm peak}>10\ \rm{km\ s}^{-1}$ and $r<300\ \rm{kpc}$ is reduced by a factor of $2/3$, while the $1\sigma$ host-to-host scatter corresponds to an $87\%$ reduction at most. This suggests that subhalo disruption due to baryonic effects, such as stellar feedback and the tidal influence of a central galactic disk, should not be neglected in semianalytic models that use the subhalo populations predicted by these DMO simulations as input. In particular, for MW-mass host halos that contain a central galactic disk similar to those found in the \texttt{m12i} and \texttt{m12f} FIRE simulations, the reduction in substructure due to the disk and other baryonic processes is larger than the scatter in subhalo abundance from host to host, so the impact of baryonic physics cannot be accounted for simply by marginalizing over the subhalo populations of host halos with a range of formation histories.

While the average amount of subhalo disruption is larger than the host-to-host scatter among the subhalo populations in these simulations, the impact of baryons on individual subhalo populations is largely consistent. In particular, our classifier predicts that the hosts with the most subhalos tend to have the largest number of surviving subhalos once baryonic effects are taken into account. Moreover, the number of DMO subhalos and the predicted number of surviving subhalos above different $V_{\rm max}$ thresholds and within various hostcentric radii are highly correlated for this simulation suite. For example, the Spearman rank correlation coefficient between the number of surviving subhalos with $V_{\rm peak}>10\ \rm{km\ s}^{-1}$ and $r<300\ \rm{kpc}$ predicted by the DMO simulations and by our classifier is $0.74$. This implies that the shapes of the velocity functions and radial distributions are not strongly affected by baryonic physics; indeed, the scaled DMO curves in Figure \ref{fig:zoomins} are very similar to the random forest predictions, except at small radii, where subhalos are preferentially disrupted in the training data. The fractional amount of subhalo disruption is also consistent among the zoom-in simulations. In particular, the number of predicted surviving subhalos with $V_{\rm max}>10\ \rm{km\ s}^{-1}$ and $r<300\ \rm{kpc}$ for all $45$ hosts is given by $N_{\rm pred}/N_{\rm DMO}=0.65\pm 0.09$. To illustrate these results, the inset in the bottom panel of Figure~\ref{fig:zoomins} shows the number of predicted surviving subhalos with $V_{\rm peak}>10\ \rm{km\ s}^{-1}$ and $r<300\ \rm{kpc}$ for each host versus the corresponding number of subhalos in each DMO simulation. The inset shows that the random forest predictions are consistent with an overall scaling of the DMO subhalo populations. Thus, subhalo disruption due to baryonic effects can be parameterized rather simply for these host halos in the context of our disruption model. We leave a detailed exploration of such a parameterization to future work informed by a wider range of hydrodynamic simulations, but we note that a simple one-parameter rescaling would not be sufficient to model subhalo disruption in detail; for example, Figure \ref{fig:zoomins} shows that the shape of the mean radial subhalo distribution is somewhat altered by baryonic physics.  Finally, we note that our random forest classifier predicts that these zoom-in simulations typically contain more high-$V_{\rm max}$ subhalos than \texttt{m12i} or \texttt{m12f} and more subhalos at small radii than \texttt{m12i}. Determining whether these differences represent statistical fluctuations or systematic differences between the FIRE simulations and this simulation suite would require a larger sample of hydrodynamic results for comparison.

\subsubsection{Implications for Satellite Searches}

Our model, when applied to MW-size zoom-in simulations, suggests that MW-mass host halos are somewhat less likely to host bright satellite galaxies such as the Magellanic Clouds and that they have more extended radial satellite profiles than those inferred from DMO simulations. At face value, both of these predictions seem to be in tension with observations of MW satellites \citep[e.g., see][]{1711.06267}. However, the MW itself could be an outlier, so here we also examine our model's predictions for the satellite populations of MW analogs.

\begin{figure}
\centering
\includegraphics[scale=0.45]{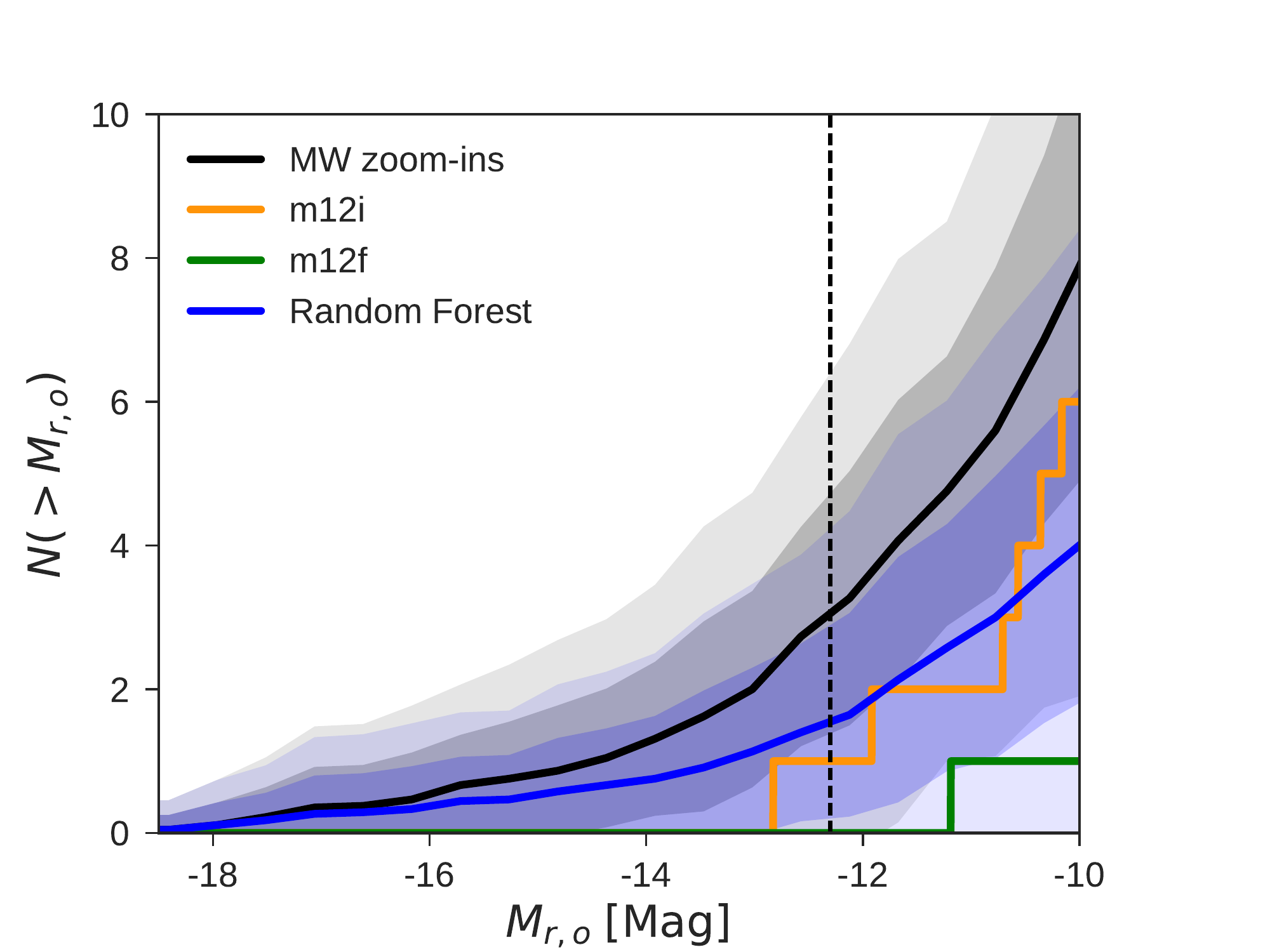}
\caption{Luminosity functions for the DMO zoom-in simulations presented in Mao et al.\ (\citeyear{Mao150302637}; black), inferred using the $V_{\rm peak}-$luminosity abundance-matching relation tuned to the $r$-band luminosity function from the GAMA survey \citep{Loveday150501003}. We do not apply scatter in the $V_{\rm peak}-M_{r,o}$ relation to highlight the host-to-host variability. The blue line shows our mean prediction for the luminosity function of surviving satellites, and shaded areas show $\pm 1\sigma$ and $\pm 2\sigma$ standard deviations. We also plot luminosity functions for \texttt{m12i} (orange) and \texttt{m12f} (green). Here $M_{r,o}$ is the observed $r$-band luminosity, and the vertical line corresponds to the completeness limit of the SAGA survey.}
\label{fig:luminosity}
\end{figure}

To estimate the impact of baryonic physics on the luminosity functions of MW, which can be compared to the results of the SAGA survey, we use the $V_{\rm peak}$-luminosity abundance-matching relation tuned to the $r$-band luminosity function from the GAMA galaxy survey \citep{Loveday150501003}; we refer the reader to Geha et al.\ (\citeyear{Mao170506743}) for details on the abundance matching procedure. Figure \ref{fig:luminosity} shows the resulting luminosity functions for the MW zoom-in suite, along with the luminosity functions for these hosts predicted by our random forest classifier. We neglect the scatter in the $V_{\rm peak}$-luminosity relation for this simple estimate because the host-to-host scatter among the zoom-in simulations is larger than the intrinsic scatter in the luminosity function introduced by abundance matching. Our classifier predicts a significant reduction in the number of bright satellites associated with MW analogs; the number of satellites with observed $r$-band magnitudes $M_{r,o}<-12.3$ inferred from the DMO simulations is $3.0\pm 1.6$, while our random forest predicts that only $1.5\pm 1.3$ such satellites exist. Although these estimates of surviving satellite populations are simplistic, it will be interesting to compare predictions informed by hydrodynamic simulations to observational results as the number of systems with high completeness limits improves.


\section{Conclusions and Discussion}
\label{Discussion}

To conclude, we summarize our main results as follows.

\begin{enumerate}
\item We train a random forest classifier on disrupted and surviving subhalos in two hydrodynamic zoom-in simulations of MW-mass host halos from the FIRE project using five properties of each subhalo: $d_{\rm peri}$, $a_{\rm peri}$, $a_{\rm acc}$, $M_{\rm acc}$, and $V_{\rm acc}$.
\item Our classifier identifies subhalos in the FIRE simulations with an $85\%$ OOB classification score and predicts surviving subhalo populations from DMO simulations of these hosts that are in excellent agreement with the hydrodynamic results, often outperforming the DMO-plus-disk simulations presented in Garrison-Kimmel et al.\ (\citeyear{Garrison-Kimmel170103792}). 
\item We argue that our classifier captures the effects of the central galactic disks that develop in the FIRE simulations, in addition to other baryonic disruption mechanisms such as stellar feedback.
\item We use our classifier to predict surviving subhalo populations for the suite of DMO zoom-in simulations of MW-mass host halos presented in Mao et al.\ (\citeyear{Mao150302637}), finding that the average amount of subhalo disruption is larger than the host-to-host scatter; however, the baryonic impact on each subhalo population is largely consistent, with $N_{\rm pred}/N_{\rm DMO}=0.65\pm 0.09$ for subhalos with $V_{\rm peak}>10\ \rm{km\ s}^{-1}$ and $r<300\ \rm{kpc}$.
\end{enumerate}
We refer the reader to the end of Section \ref{RandomForest} for a summary of the limitations of our classification method.

There are several interesting avenues for future work. For example, since we find that the average amount of subhalo disruption due to baryonic physics is larger than the host-to-host scatter among the suite of zoom-in simulations analyzed above, this characteristic reduction in the number of subhalos should be taken into account when marginalizing over the effects of baryonic physics for MW-mass host halos that contain a central galactic disk. Thus, it is plausible that the reduced number of surviving subhalos will change the conclusions drawn from semianalytic models that use the subhalo populations predicted by such simulations (e.g., \citealp{Lu160502075,Lu170307467}).

Another potential application of our results concerns the radial segregation of dark matter subhalos with respect to various subhalo properties. Subhalo segregation, as studied by van den Bosch et al.\ (\citeyear{VandenBosch151001586}), directly depends on the subhalo populations predicted by DMO simulations. Since subhalo populations that have been altered by baryonic effects systematically differ from those predicted by DMO simulations, subhalo segregation could be affected by baryonic physics, and our classifier provides an efficient method for predicting surviving subhalo populations in order to explore this possibility. Meanwhile, resolving the disruption of individual subhalos in detail is an important challenge for current simulations; for example, van den Bosch (\citeyear{VandenBosch161102657}) estimated that $80\%$ of all subhalo disruption in the Bolshoi simulation is numerical, rather than physical. The Latte simulations have $\sim 4000\times$ smaller dark matter particle mass and $\sim 35\times$ smaller dark matter force softening than Bolshoi, so these effects are likely much less severe, but it is nevertheless worth exploring whether artificial disruption persists in high-resolution hydrodynamic simulations and how these numerical artifacts might influence our results.

The algorithm presented in this paper is extremely simple, using only five subhalo properties as training features. Nevertheless, these properties encode the majority of the information about subhalo disruption in the \texttt{m12i} and \texttt{m12f} FIRE simulations, yielding a classifier that predicts surviving subhalo populations from DMO simulations that are in excellent agreement with hydrodynamic results. Of course, as the number of hydrodynamic training simulations grows, it will be worthwhile to explore more sophisticated classification algorithms and to study the feature selection in more detail. It will be interesting to assess how well a classifier can perform in principle, since there are characteristic differences between DMO and hydrodynamic simulations, including changes in subhalo orbits due to the presence of baryons~\citep{Zhu170105933}, that our simple model cannot capture. Our results hint that these characteristic differences are relatively unimportant, but further tests should be performed using a larger sample of training simulations.

As more high-resolution zoom-in simulations become available, it will become feasible to train classifiers on increasingly diverse datasets, allowing for more robust predictions. Once a classifier has been trained on a wide variety of hydrodynamic simulations, it can predict a range of surviving subhalo populations associated with different central galaxy types and halo formation histories directly from DMO simulations. It is worth exploring whether these predictions can be used as input for neural networks in order to generate large samples of mock halo catalogs, perhaps eliminating the need for certain types of simulations entirely.

Machine-learning algorithms have the potential to identify large samples of realistic subhalo populations that can be used as input for models that populate subhalos with galaxies. Comparing the surviving subhalo populations predicted by such algorithms for host halos on different mass scales could provide insight into the original TBTF problem for MW-mass systems and into analogous problems for host halos of different masses. Moreover, comparing the results of classification algorithms that are trained on hydrodynamic simulations with different implementations of baryonic physics would be a promising step toward parameterizing the impact of baryons on the abundance and properties of dark matter subhalos.


\acknowledgments
We have made our code and trained classifier publicly available at \href{https://github.com/ollienad/subhalo\_randomforest}{github.com/ollienad/subhalo\_randomforest}; please contact the authors with data requests. We thank Frank \mbox{van den Bosch} and Andrew Hearin for useful discussions.  This research was supported in part by NSF grant AST-1517148. YY-M is supported by the Samuel P.\ Langley PITT PACC Postdoctoral Fellowship.
Support for SG-K was provided by NASA through Einstein Postdoctoral Fellowship grant number PF5-160136 awarded by the Chandra X-ray Center, which is operated by the Smithsonian Astrophysical Observatory for NASA under contract NAS8-03060. AW was supported by a Caltech-Carnegie Fellowship, in part through the Moore Center for Theoretical Cosmology and Physics at Caltech, and by NASA through grants HST-GO-14734 and HST-AR-15057 from STScI.

This research made use of computational resources at SLAC National Accelerator Laboratory, a U.S.\ Department of Energy Office; the authors are thankful for the support of the SLAC computational team.  
This research was supported in part by the National Science
Foundation under grant No.\ NSF PHY17-48958 through the Kavli Institute for
Theoretical Physics program ``The Galaxy-Halo Connection
Across Cosmic Time.'' 
This research made use of the Python Programming Language, along with many community-developed or maintained software packages, including
IPython \citep{ipython},
Jupyter (\http{jupyter.org}),
Matplotlib \citep{matplotlib},
NumPy \citep{numpy},
Pandas \citep{pandas},
\texttt{Scikit-Learn} \citep{scikit-learn}, SciPy \citep{scipy}, and Seaborn (\http{seaborn.pydata.org}).
This research made extensive use of the {\tt arXiv} and NASA's Astrophysics Data System for bibliographic information.

\bibliographystyle{yahapj}
\bibliography{references,software}


\appendix
\section{Scatter, Feature Selection, and Resolution}
\label{appendix}

\begin{figure*}[b]
\centering
\includegraphics[scale=0.46]{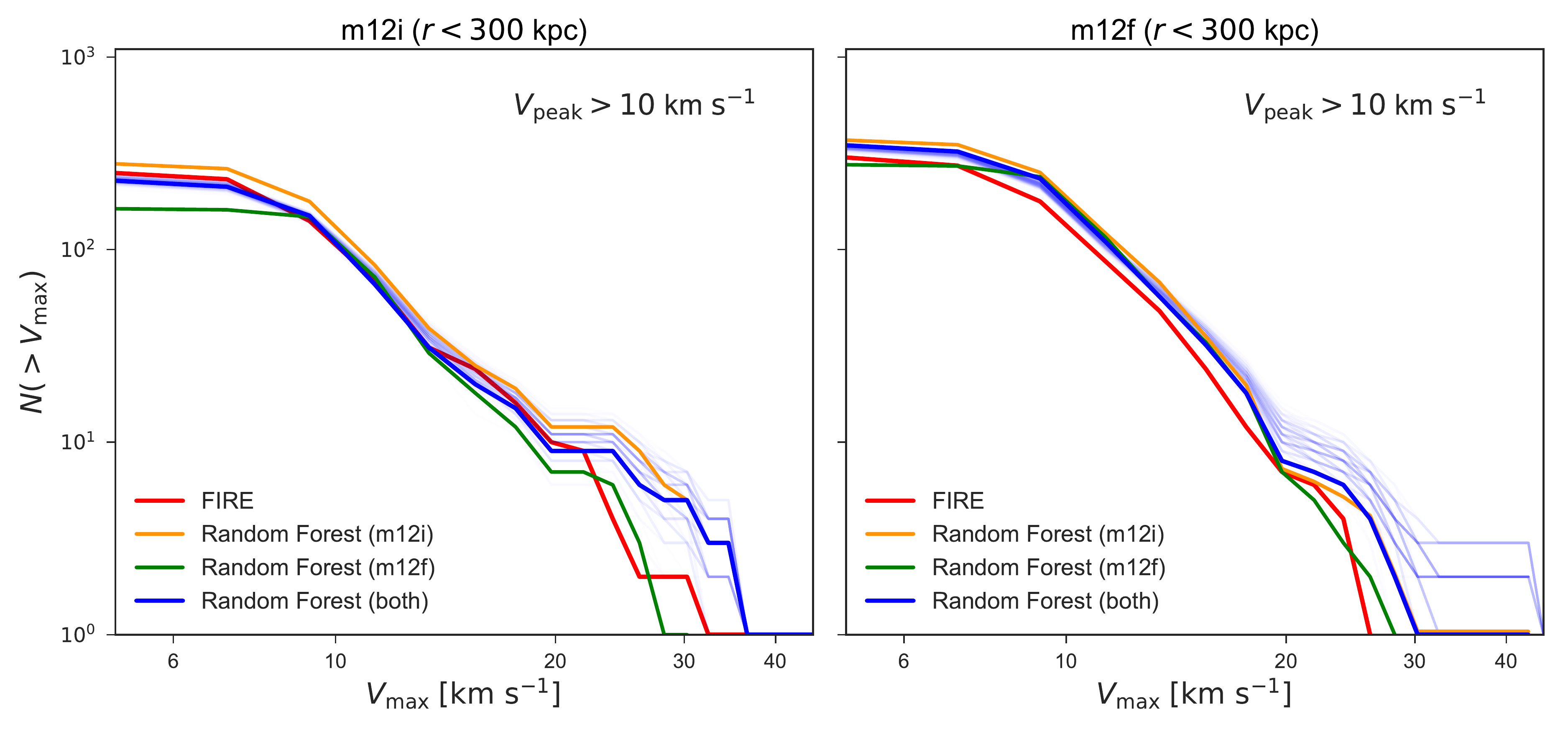}
\caption{Velocity functions for \texttt{m12i} (left) and \texttt{m12f} (right) predicted by the most probable realization of our random forest classifier when trained only on subhalos from \texttt{m12i} (orange) or \texttt{m12f} (green) with $V_{\rm peak}>10\ \rm{km\ s}^{-1}$. Blue lines show $200$ realizations of the prediction for our fiducial classifier, which is trained on subhalos from both hosts, and red lines show the FIRE results. While there is a difference between the total number of surviving subhalos predicted by classifiers trained only on \texttt{m12i} or \texttt{m12f}, the scatter about the most probable prediction for our fiducial classifier is small.}
\label{fig:variance}
\end{figure*}

We perform several tests to check the robustness of our results. First, we examine the scatter in the random forest predictions for the \texttt{m12i} and \texttt{m12f} $V_{\rm{max}}$ functions and whether the details of the training data affect our results. Figure \ref{fig:variance} shows the $V_{\rm{max}}$ functions from $200$ realizations of our fiducial classifier, along with the most probable realization of classifiers trained only on subhalos from \texttt{m12i} or \texttt{m12f} with $V_{\rm peak}>10\ \rm{km\ s}^{-1}$. The scatter about the most probable prediction for our fiducial classifier is small; in particular, the intrinsic scatter of the random forest prediction is comparable to or smaller than the Poisson noise over the entire velocity function for each host. Thus, even though the prediction for the total number of surviving subhalos is different for classifiers trained on \texttt{m12i} or \texttt{m12f} separately, this uncertainty does not propagate to our fiducial classifier.

\begin{figure*}
\centering
\includegraphics[scale=0.46]{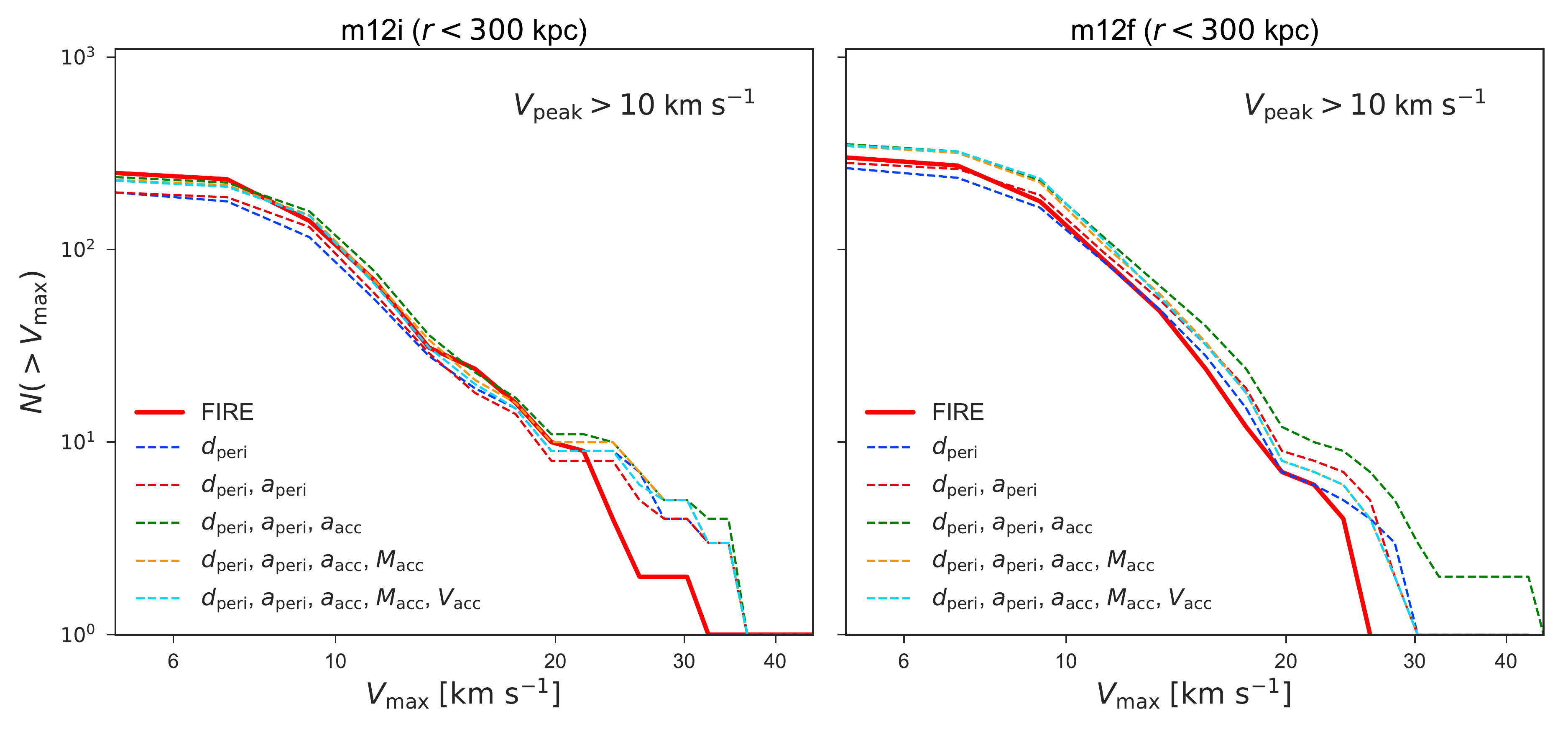}\\
\includegraphics[scale=0.46]{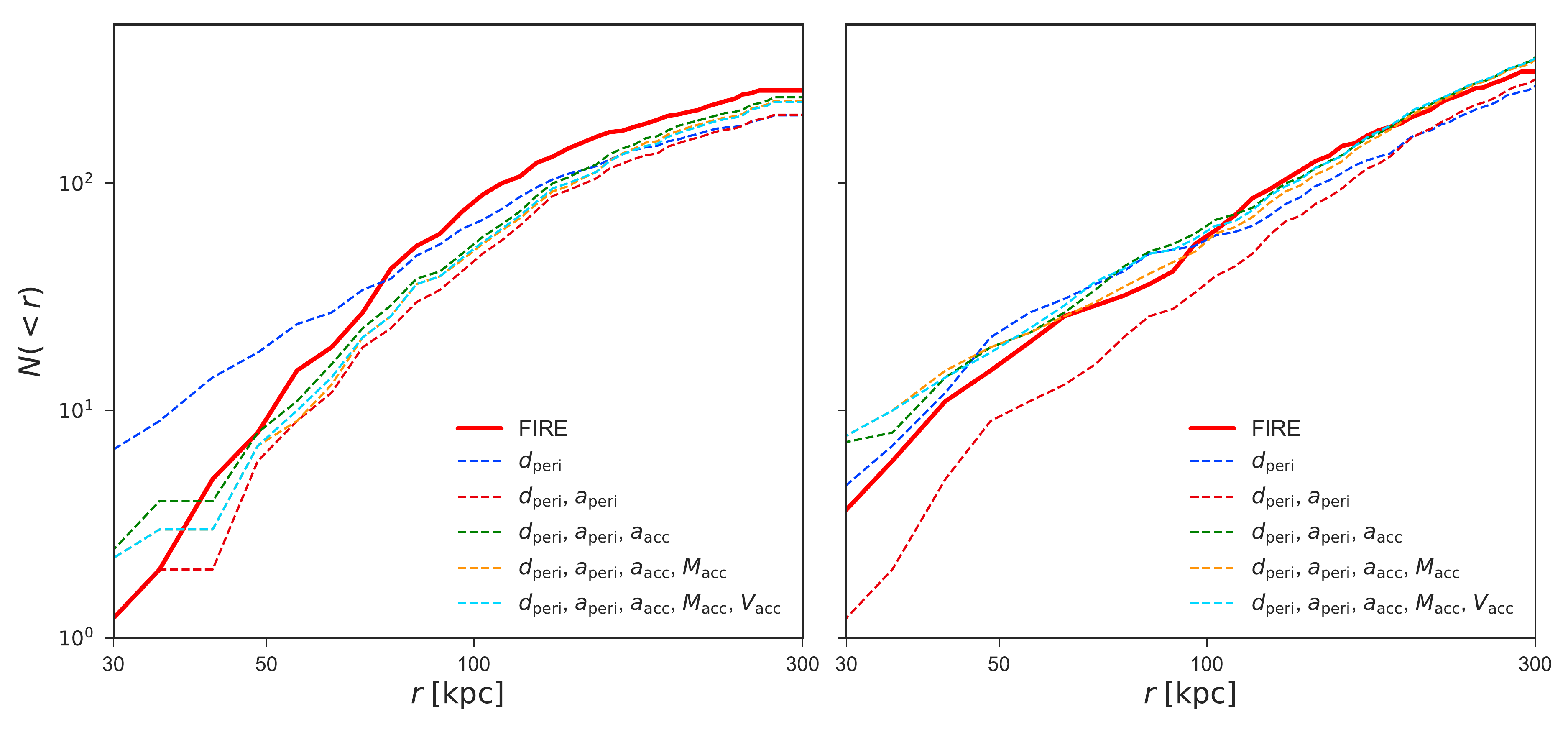}
\caption{Velocity functions and radial distributions of subhalos in \texttt{m12i} (left) and \texttt{m12f} (right) predicted by the most probable realization of random forest classifiers trained on subhalos from both hydrodynamic simulations with $V_{\rm peak}>10\ \rm{km\ s}^{-1}$. The classifiers use the features $d_{\rm peri}$ (blue); $d_{\rm peri}$ and $a_{\rm peri}$ (red); $d_{\rm peri}$, $a_{\rm peri}$, and $a_{\rm acc}$ (green); $d_{\rm peri}$, $a_{\rm peri}$, $a_{\rm acc}$, and $M_{\rm acc}$ (orange); and $d_{\rm peri}$, $a_{\rm peri}$, $a_{\rm acc}$, $M_{\rm acc}$, and $V_{\rm acc}$ (cyan), corresponding to the rows of Table \ref{tab:percent} and the columns of Figure \ref{fig:barplot}. The solid red lines show the FIRE results.}
\label{fig:features}
\end{figure*}

Next, we explore the choice of training features. In particular, we test how adding subhalo features affects our results for the $V_{\rm{max}}$ functions and radial distributions of the surviving subhalo populations predicted from DMO simulations of \texttt{m12i} and \texttt{m12f}. In Figure \ref{fig:features}, we plot the most probable $V_{\rm max}$ functions and radial distributions predicted by five classifiers that each use an additional training feature, corresponding to the rows of Table \ref{tab:percent} and the columns of Figure \ref{fig:barplot}. As we add subhalo features, the predicted distributions approach the FIRE results. Interestingly, $d_{\rm peri}$ (or $a_{\text{peri}}$ or $a_{\text{acc}}$) alone provides most of the information needed to match the total number of surviving subhalos with $V_{\rm peak}>10\ \rm{km\ s}^{-1}$ and $r<300\ \rm{kpc}$, but adding additional features improves the predictions at large $V_{\rm max}$ and small radii.

\begin{figure*}
\centering
\includegraphics[scale=0.46]{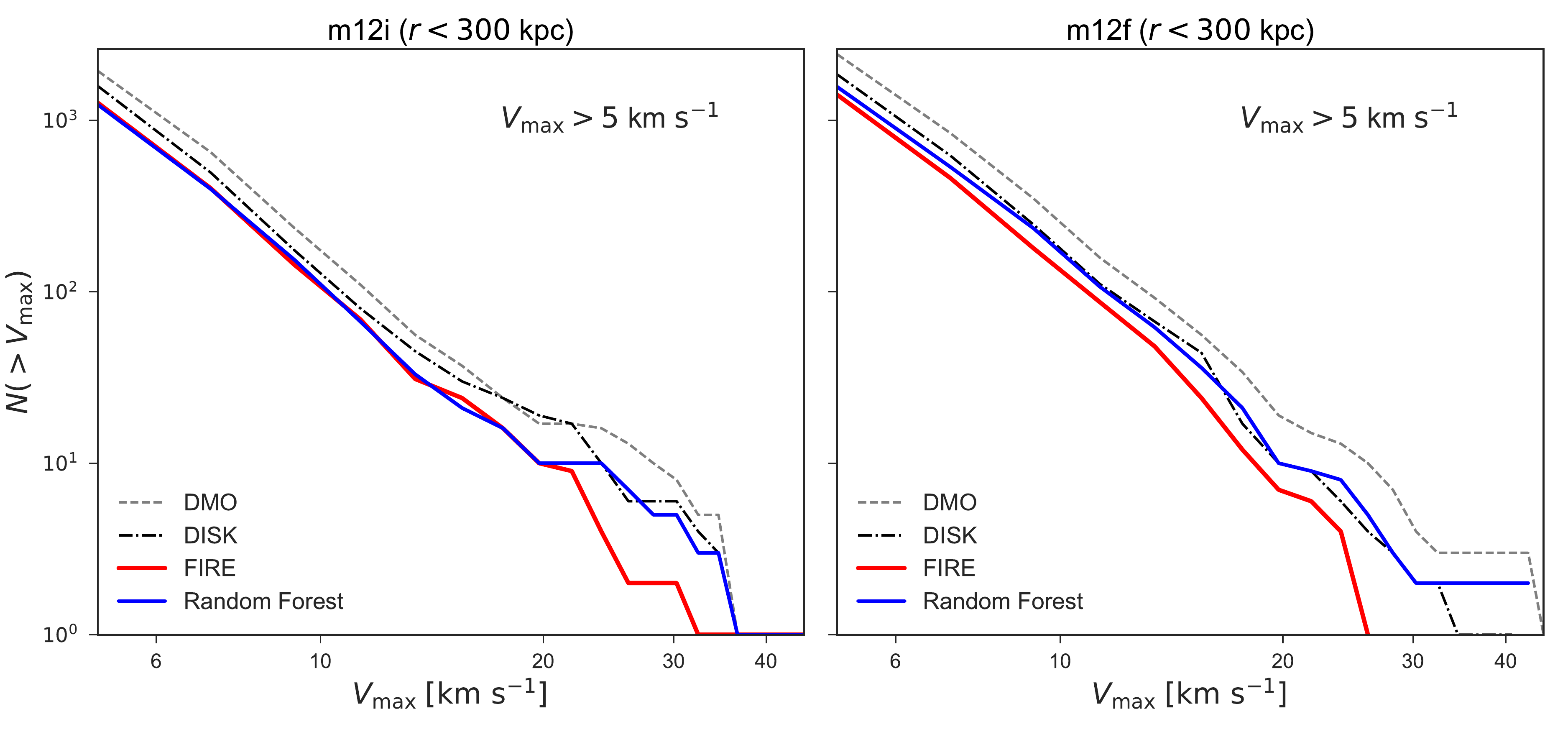}\\
\includegraphics[scale=0.46]{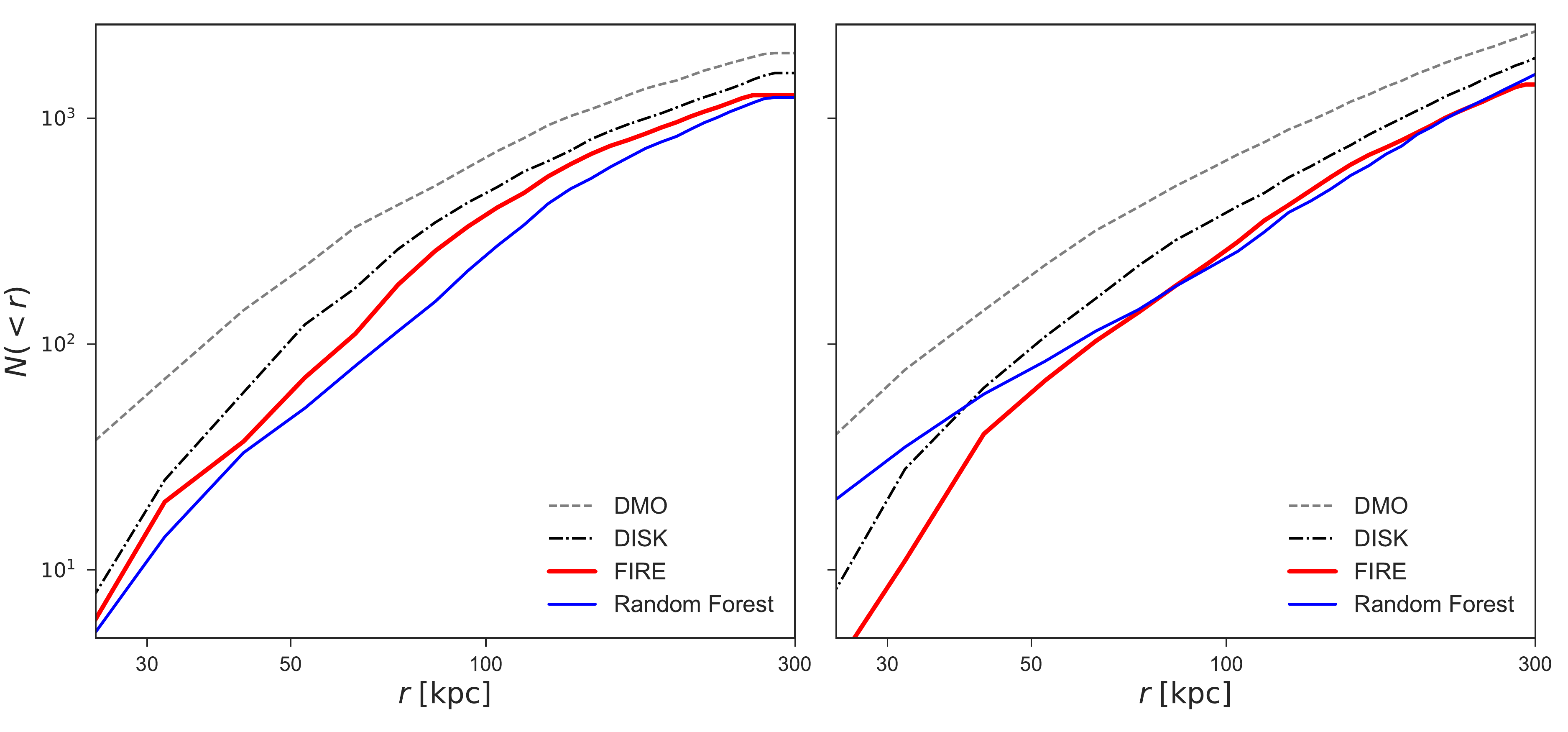}
\caption{Velocity functions (top) and radial distributions (bottom) for subhalos hosted by \texttt{m12i} (left) and \texttt{m12f} (right), predicted by the most probable realization of our random forest classifier trained on subhalos from \texttt{m12i} and \texttt{m12f} with $V_{\rm peak}>5\ \rm{km\ s}^{-1}$ (blue). The FIRE (red), DISK (dot-dashed), and DMO (dashed) results are shown for comparison. We restrict these plots to subhalos within $300\ \rm{kpc}$ of their respective hosts at $z=0$ and with $V_{\rm max}>5\ \rm{km\ s}^{-1}$, where $V_{\rm max}$ is the maximum circular velocity evaluated at $z=0$, to allow for a direct comparison with the results in \citet{Garrison-Kimmel170103792}.}
\label{fig:comparison}
\end{figure*}

Finally, we study how our results depend on the resolution limits used for the training data. In Figure \ref{fig:comparison}, we show the \texttt{m12i} and \texttt{m12f} velocity functions and radial distributions predicted by a classifier trained on subhalos from both hosts with $V_{\rm peak}>5\ \rm{km\ s}^{-1}$, which is less restrictive than the $V_{\rm peak}>10\ \rm{km\ s}^{-1}$  cut used in our primary analysis. We plot the results for subhalos with $V_{\rm max}>5\ \rm{km\ s}^{-1}$, where $V_{\rm max}$ is the maximum circular velocity evaluated at $z=0$, which allows for a direct comparison with the results in \citetalias{Garrison-Kimmel170103792}. Our conclusions are unaffected by changing the minimum circular velocity. In fact, our predictions match the hydrodynamic results even more closely than before in the low-$V_{\rm max}$ regime, since this less restrictive cut significantly increases the number of subhalos at the low-$V_{\rm max}$ end of the training set. Thus, our classifier can be applied to simulations with a range of resolution thresholds if appropriate cuts are applied to the training data.

\end{document}